\documentclass{article}
\usepackage[utf8]{inputenc}
\usepackage[left=3cm,
            right=2.5cm,
            top=2.5cm,
            bottom=3cm
            ]{geometry}  

\usepackage{tcolorbox}
\usepackage{amsmath,amsfonts,amssymb,array,graphicx,color}
\usepackage[dont-mess-around]{fnpct}

\usepackage{blkarray} 
\usepackage{comment}
\usepackage{authblk}
\usepackage{appendix}

\usepackage{hyperref}
\hypersetup{colorlinks,bookmarksopen,bookmarksnumbered,
 citecolor=magenta,
 linkcolor=blue}
 \usepackage[style=phys,hyperref=true,articletitle=true,biblabel=brackets,chaptertitle=false]{biblatex}
\numberwithin{equation}{section}

\usepackage{braket}  
\def \be {\begin{equation}} 
\def \ee {\end{equation}} 
\def \l {\left(} 
\def \r {\right)} 
\def \la {\langle} 
\def \ra {\rangle}  

\date{}

\title{
Thermal entanglement in conformal junctions
}
\author{Luca Capizzi$^{1,2}$, Andrei Rotaru$^{1,3}$}
\date{}

\begin{document}

\maketitle
$^1$SISSA and INFN Sezione di Trieste, via Bonomea 265, 34136 Trieste, Italy.\\

$^2$Universit\'e Paris-Saclay, CNRS, LPTMS, 91405, Orsay, France.\\

$^3$Sorbonne Universit\'e, CNRS, Laboratoire de Physique Th\'eorique et Hautes \'Energies, LPTHE, F-75005 Paris, France

\begin{abstract}
We consider a quantum junction described by a 1+1-dimensional boundary conformal field theory (BCFT). Our analysis focuses on correlations emerging at finite temperature, achieved through the computation of entanglement measures. Our approach relies on characterizing correlation functions of twist fields using BCFT techniques. We provide non-perturbative predictions for the crossover between low and high temperatures. An intriguing interplay between bulk and boundary effects, associated with the bulk/boundary scaling dimensions of the fields above, is found. In particular, the entanglement entropy is primarily influenced by bulk thermal fluctuations, exhibiting extensiveness for large system sizes with a prefactor independent of the scattering properties of the defect. In contrast, negativity is governed by fluctuations across the entangling points only, adhering to an area law; its value depends non-trivially on the defect, and it diverges logarithmically as the temperature is decreased. To validate our predictions, we numerically check them for free fermions on the lattice, finding good agreement.
\end{abstract}

\section{Introduction}

Single localized defects may alter macroscopically the behaviour of a quantum system. For one-dimensional critical quantum many-body systems, these effects are particularly enhanced, due to the presence of slow algebraic decay of correlations that propagate from the defect to the boundary. As a result, many counterintuitive phenomena have been observed in those systems, such as the famous Kondo effect \cite{Kondo-64}.

 In the context of Luttinger liquids, it has been understood by Kane and Fisher \cite{kf-92}, that, a single defect can be relevant, marginal or irrelevant. In the relevant case, which is particularly important for the Kondo Physics, the defect screens the correlation across it, and, in the infrared regime, the two halves of the systems become effectively decoupled. The underlying explanation for this phenomenon is nowadays formalized in the language of Boundary Conformal Field Theory (BCFT) \cite{al-91,al-93,oa-96}.
 
However, residual quantum correlations are still present at finite temperatures. Their quantitative understanding had been a challenge for a long time, but a recent work \cite{kss-21} shed light on their origin. In particular, the logarithmic negativity, introduced in Ref. \cite{Plenio-05} as a good measure of entanglement in mixed states, has been tackled in Ref. \cite{kss-21} via a non-trivial perturbation scheme around the infrared fixed point.

Marginal defects are even more puzzling. In this case, the perturbation induced by the defect is invariant under the renormalization group, and novel infrared fixed points are found, corresponding to partially permeable interfaces. It has been shown that the ground state entanglement of a half chain attached to the defect grows logarithmically with the system size - with a universal prefactor which depends explicitly on the scattering properties of the defect \cite{Peschel-05,ss-08}. Since the prefactor above is related to the central charge for homogeneous system \cite{cc-04,cc-09}, some literature has referred to an \textit{effective central charge} in the presence of the interface. Besides this vague analogy, this terminology is highly misleading, since the bulk theory does not depend on the boundary condition, and, for instance, the associated central charge is not affected by the properties of the defect. Analytical investigation of the ground state entanglement for free systems (fermions and bosons) in the presence of interfaces has been pursued via Conformal Field Theory (CFT) techniques in Refs. \cite{gm-17,cmc-22,cmc-22a}, with lattice/microscopic methods in \cite{ep-10,isl-09,cmv-12}, and via the modular Hamiltonian in Ref. \cite{Mintchev:2020jhc}. We also mention that linear growth of entanglement entropy after global quenches has been found in Refs. \cite{wwr-17,ce-23}.

While extensive studies of thermal entanglement in CFT are present in the literature, see for example Refs. \cite{hn-13,dd-14,cw-14,ch-14,hn-15,cct-14,cw-15,hs-16,sr-19,wlckg-20,rmc-23}, and boundary effects have been thoroughly characterized \cite{bs-06,als-09,ntu-12,Takayanagi-11,bs-16,eir-22,eir-23,eirt-23}, few results are present for conformal defects.  Importantly, a CFT approach to addressing this problem has been put forward in the work of Sakai and Satoh \cite{ss-08}, in which the computation of the entropy is equivalently formulated as a partition function with the insertion of interface operators along the spatial direction. This approach, which has been used to characterize the ground state entropy between wires of the junction, turned out to be difficult to generalize to tackle other spatial bipartitions, thermal effects, excited states, dynamics, and so on.

A solution to these issues has been proposed in \cite{ce-23}, where the junction is formulated as a boundary CFT, and the entanglement is computed as a correlation function of certain twist fields. The novelty is that a family of twist fields, with a non-trivial boundary scaling dimension, has to be considered. For instance, these boundary scaling dimensions are precisely at the core of the unusual logarithmic growth of the ground state entropy, and they are related to the effective central charges observed in previous works.

\begin{figure}[h]
    \centering
    \includegraphics[scale=2]{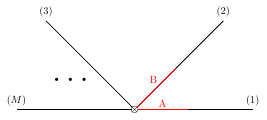}
    \caption{A junction with $M$ wires. A tripartion $A\cup B\cup \overline{(A\cup B)}$, with $A$ and $B$ intervals attached to the defect, is shown.}
    \label{fig:wires_intervals}
\end{figure}

The aim of this work is to provide a qualitative and quantitative understanding of the thermal entanglement of conformal junctions (of $M$ wires), corresponding to marginal defects, via BCFT techniques, using the formalism of bulk/boundary twist fields. In particular, we characterize both the mutual information and the logarithmic negativity of spatial regions, as depicted in Fig. \ref{fig:wires_intervals}. 
To the best of our knowledge, both our approach and the results we found are new. For instance, the emerging physics strongly differs from the behaviors observed in Ref. \cite{kss-21} for relevant defects. 

In our analysis, we are able to get general closed analytical formulae describing the exact crossover as functions of the temperature. We also describe methods to tackle systematically the high/low-temperature regimes, using quantization over the direct/crossed channels respectively.
Specifically, the sole required input consists of the boundary scaling dimension of the twist fields, which depends on the theory and the underlying defect. These dimensions correspond to the established prefactor of the logarithmic scaling of the ground state entropy, and we provide a relationship between their values and the entanglement measures by exploiting the conformal symmetry of the model.
We find that the entropies are, generically extensive with the subsystem size, as long as the typical geometrical lengths are much larger compared to the thermal length $\beta$, and the density of entropy does not depend on the details of the defect. This is physically reasonable since for the case considered above the dominant contribution to the entropy is due to classical correlations between the subsystem and a fictitious thermal bath. In contrast, both the mutual information and the logarithmic negativity of intervals attached to the defect satisfy an area law: the saturation value diverges logarithmically as the temperature decreases, and the universal prefactor is related to the boundary scaling dimensions of twist fields mentioned above. We finally mention that the same quantities remain finite for the relevant defect of Ref. \cite{kss-21}, and they do not show any logarithmic divergence.

We organize our manuscript as follows. In Sec. \ref{sec:CFT_half_line} we provide a general approach to compute correlation functions of twist fields at finite temperature in the presence of a single boundary point. At this stage, we do not specify the theory nor the fields, and most of our results will be consequences of conformal invariance, modularity, and Operator Product Expansion (OPE). We generalize the approach to tackle finite-size geometries (with some limitations) in Sec. \ref{sec:CFT_finite_size}, and we discuss some effects arising from the discrete (point) spectrum of the Hamiltonian. We use those results to give analytical predictions for the entropy and the negativity in Sec. \ref{sec:Ent_measures}. We finally cross-check numerically some results in 
 Sec. \ref{sec:Numerics} for a free fermion chain with a conformal defect. Conclusions and outlook are left to Sec. \ref{sec:Conclusions}.

\section{CFT approach: thermal effects on the half-line}\label{sec:CFT_half_line}

In this section, we give a field theoretical description of a conformal junction of $M$ wires at finite temperature. We consider semi-infinite extended wires attached to a single defect, and we represent it as a theory over the half-plane with multiple species of fields and central charge $c$ - a procedure known in the literature as the \textit{folded picture} \cite{gm-17}. Since we work at finite temperature $\beta^{-1}$, the underlying Euclidean geometry is the half-infinite cylinder
\be
\text{Re}\l \zeta\r \in [0,\infty), \quad \text{Im}(\zeta) \in [0,\beta).
\ee
Periodic boundary conditions are chosen along the (Euclidean) time direction, while boundary conditions of type $b$, corresponding to the scattering properties of the defect, are kept along $\text{Re}\l \zeta\r =0$. We assume that $b$ is conformal-invariant, and information about the boundary condition is encoded in a \textit{conformal boundary state} $\ket{b}$ \cite{Cardy-84}. Such an objected is generically constructed using states from the Hilbert space of the corresponding bulk system, in our case a ring of length $\beta$ evolving in real time. We refer to this choice of quantization as the "crossed" channel.

Our goal is to characterize the multi-point correlation functions of twist fields in the aforementioned geometry. These quantities play a crucial role in expressing various entanglement measures between spatial regions \cite{cc-09}. In principle, a theory contains numerous twist fields, each associated with different monodromy conditions with respect to the local fields \cite{bd-17,Dupic:2017hpb}. However, the forthcoming discussion is general and applies to any primary twist field, namely as long as it is the lightest and, therefore, the most relevant operator compatible with these specific semilocal properties.

 Let us consider a  twist field $\sigma(\zeta,\bar{\zeta})$inserted in the bulk and let $\tilde{\sigma}(\zeta,\bar{\zeta})$ be its hermitian conjugate. We assume that $\sigma(\zeta,\bar{\zeta})$
 is a primary field with a given scaling dimension $\Delta$. In contrast, we refer to $\sigma_b$ as the corresponding boundary field arising from the leading term in the bulk-to-boundary OPE \cite{Cardy-04} expansion generated by the $\sigma$ operator
\footnote{Compared to the $\beta=\infty$ case for which the relevant BCFT geometry is the half-plane, at finite temperature the boundary has non-zero intrinsic curvature. This requires the modification of the bulk-boundary OPE as explained in \cite{runkel2000boundary}. Conveniently, the leading term in  \eqref{eq:bulk_boundary_OPE_cyl} receives no curvature corrections, for $\sigma_b$ is primary. Also, the proportionality factors, known in the BCFT literature as bulk-boundary structure constants, have been discarded and absorbed in the normalization of the fields. In some contexts, they are nevertheless important, and, for example, they can be related to the Affleck-Ludwig boundary entropy \cite{al-91a}. However, they do not play a role in our discussion, as we will consider entanglement measures up to additive constants.}
\be
\label{eq:bulk_boundary_OPE_cyl}
\sigma(\zeta,\bar{\zeta})\simeq \frac{1}{\zeta^{\Delta-\Delta_b}}\sigma_b(0)+\dots, \quad \zeta \rightarrow 0.
\ee

\begin{figure}[t]
    \centering
\includegraphics[width=0.6\linewidth]{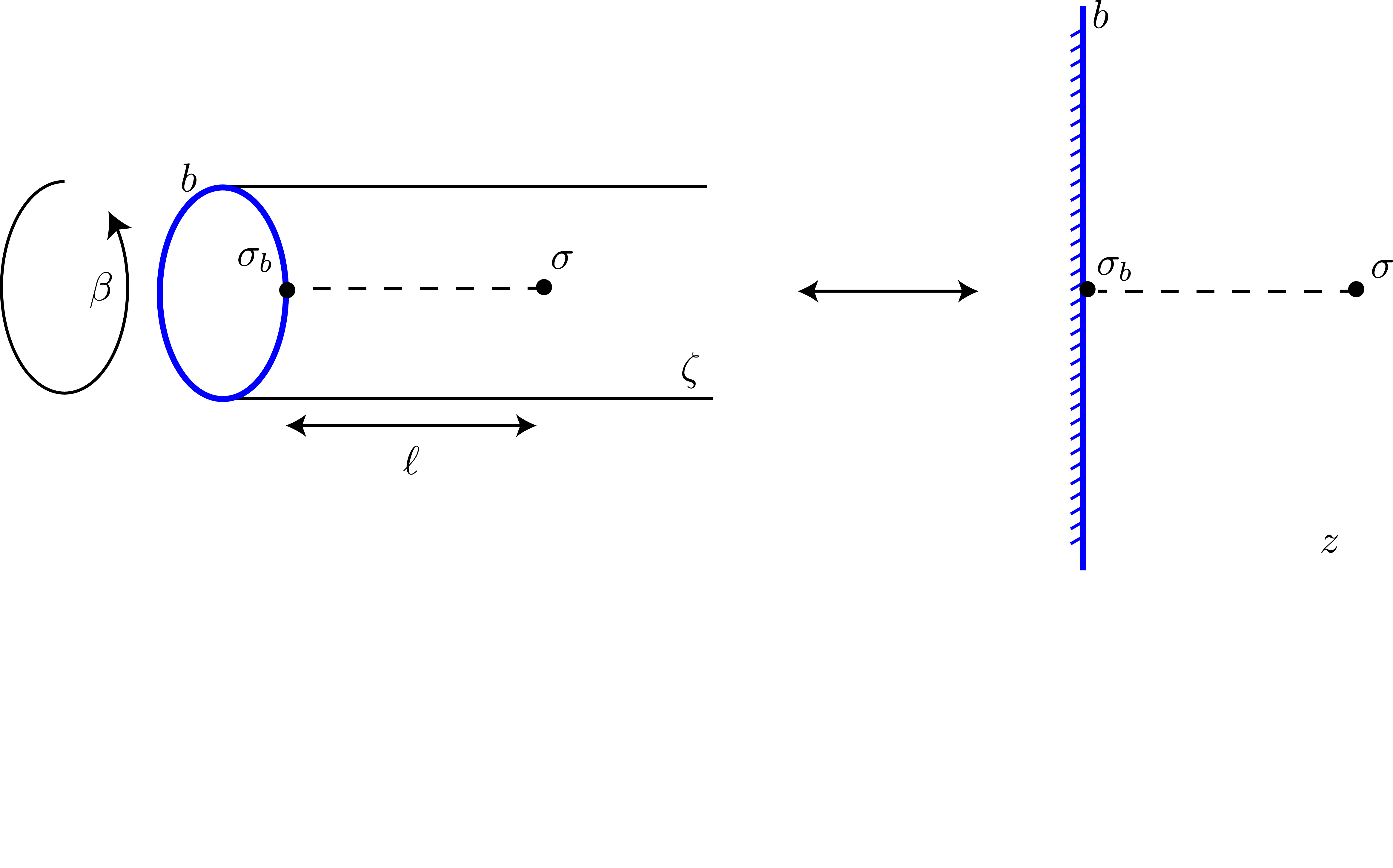}
    \caption{Two-point function of twist fields. Through the transformation $\zeta \rightarrow z(\zeta)$ \eqref{eq:zzeta_transf}, the half-infinite cylinder is mapped onto the half-plane $\text{Re}(z)\geq 0$. The blue line, that is the boundary of the geometry, encodes the properties of the defect as a boundary condition $b$.}
    \label{fig:Transf}
\end{figure}

Here $\Delta_b$ is the (boundary) scaling dimension of $\sigma_b$ which, in general, depends on the chosen boundary conditions $b$. While the precise computation of a generic multipoint function requires in principle a finer characterization of the BCFT under analysis, some salient features depend on the bulk/boundary scaling dimensions $\Delta,\Delta_b$ only. In particular, as we show below, the two-point function relevant for computing moments of the RDM for intervals adjacent to the boundary
\be
\la \sigma_b(0)\tilde{\sigma}(\zeta,\bar{\zeta})\ra, \quad\quad \text{Im}(\zeta) = 0,
\ee
  is fixed by global conformal invariance. To see it, we first apply the conformal transformation
\be
w = \exp\l \frac{2\pi \zeta}{\beta}\r,
\ee
which maps the semi-infinite cylinder onto $|w|\geq 1$. Subsequently, we utilize the transformation
\be\label{eq:zzeta_transf}
z = \frac{w-1}{w+1} = \tanh \left(\frac{\pi\zeta}{\beta}\right),
\ee
which brings us to the half-plane where $\text{Re}(z)\geq 0$.
In the $z$ geometry, the correlation function is fixed by scale invariance, and we get
\be\label{eq:corr_sigma_z}
\la \sigma_b(0)\tilde{\sigma}(z,\bar{z})\ra = \frac{1}{z^{\Delta+\Delta_b}}, \quad z \in (0,+\infty),
\ee
where the (unit) proportionality constant has been absorbed in the normalization of the  $\sigma_b$ field. Then, we employ the transformation law of primary fields \cite{dms-97} and we get
\be
\la \sigma_b(1)\tilde{\sigma}(w,\bar{w})\ra = \big|\frac{dz_1}{dw_1}\big|^{\Delta_b}\big|\frac{dz}{dw}\big|^{\Delta} \la \sigma_b(0)\tilde{\sigma}(z,\bar{z})\ra = \frac{2^{\Delta-\Delta_b}}{(w+1)^{\Delta-\Delta_b}(w-1)^{\Delta+\Delta_b}},
\ee
where we have evaluated the previous expression for $z_1 = 0$ ($w_1=1$). Finally, we go back to the initial geometry, and we obtain
\be \label{eq:corr_sigma_zeta}
\la \sigma_b(0)\tilde{\sigma}(\zeta,\bar{\zeta})\ra = \big|\frac{dw_1}{d\zeta_1}\big|^{\Delta_b}\big|\frac{dw}{d\zeta}\big|^{\Delta} \la \sigma_b(1)\tilde{\sigma}(w,\bar{w})\ra = \l\frac{\pi}{\beta}\r^{\Delta+\Delta_b} \frac{1}{\cosh\l \frac{\pi \zeta}{\beta}\r^{\Delta-\Delta_b} \sinh\l \frac{\pi \zeta}{\beta}\r^{\Delta+\Delta_b}},
\ee
that is the main result of this section. Remarkably, a close universal function depending on the dimensionless parameter $\zeta/\beta$ is achieved, and it describes the exact temperature crossover. We represent the construction above in Fig. \ref{fig:Transf}.

Clearly, the two-point function considered above is very special, and other correlation functions depend explicitly on more specialised data of the CFT under analysis. However, some general properties are expected to be independent of the underlying model. To show that, in the next subsections we recover with different techniques the asymptotic behavior of Eq. \eqref{eq:corr_sigma_zeta} in the regimes of high and low temperature. These approaches are, in principle, generalizable to more complicated geometries, and they shed light on the salient features of the two regimes mentioned above. 

\subsection{Low-temperature limit}
At finite temperatures, a correlation length $\sim \beta$ is present. Here we consider the  regime of distances much smaller than $\beta$. For instance, let us take the limit $\zeta/\beta \ll 1$ of Eq. \eqref{eq:corr_sigma_zeta}
\be\label{eq:corr_sigma_zeta_low}
\la \sigma_b(0)\tilde{\sigma}(\zeta,\bar{\zeta})\ra \simeq \frac{1}{\zeta^{\Delta+\Delta_b}}\l 1-\l \frac{\pi \zeta}{\beta}\r^2 \l  \frac{2}{3}\Delta-\frac{1}{3} \Delta_b\r+ \dots\r.
\ee
The first term gives the same divergence encountered in Eq. \eqref{eq:corr_sigma_z}, and it just corresponds to the zero-temperature expectation value of the two-point function above. The second one is an algebraic correction of order $\sim \l\frac{\zeta}{\beta}\r^2$, and it gives the first non-trivial thermal correction.

One way to understand the mechanism above is via OPE product expansion. Namely, one has to analyze the bulk-to-boundary OPE arising from the collision between $\sigma_b$ and $\tilde{\sigma}$, keeping the most relevant operators. In this limit, the branch-cut extending between the two twist fields disappears, and local boundary fields are generated. The leading contribution comes clearly from the identity field, arising from the fusion
\be
\sigma_b \times \tilde{\sigma} \rightarrow 1.
\ee
Other boundary primary fields are generated as well in the OPE, but their expectation value on the thermal state vanishes, thus they do not contribute to Eq. \eqref{eq:corr_sigma_zeta_low}. The correction $\sim \l\frac{\zeta}{\beta}\r^2$ has to be related to a (boundary) descendant field of scaling dimension $2$: this field is precisely the boundary stress-energy tensor, denoted by $T_b$, whose expectation value is
\be
\la T_b(\zeta = 0)\ra \sim \frac{1}{\beta^2},
\ee
up to a dimensionless proportionality constant. Similarly, the higher corrections appearing in Eq. \eqref{eq:corr_sigma_zeta_low} are expected to be related to the other descendants in the tower of the identity.

While the OPE approach is predictive, and it is compatible with the result \eqref{eq:corr_sigma_zeta_low}, one could \textit{naively} follow another strategy, expressing the thermal expectation value via an expansion over the first excited states above the ground state. However, this is not particularly helpful, since in the strict infinite-volume limit we are considering, there is no energy gap above the ground state, and the spectrum of the Hamiltonian $H$ is continuous\footnote{Strictly speaking, a continuous spectrum means that no (normalizable) eigenstates are present above the ground-state.}. One way to cure the issue is to put the system in a box of length $L$, with $L\gg \beta\gg \zeta$. In this way, the low-energy eigenstates have a small energy $E \sim 1/L$, but many of them still contribute non-trivially to the thermal expectation value. We will come back to this approach in Sec. \eqref{sec:CFT_finite_size}.

\subsection{High-temperature limit}\label{sec:halfline_highT}

The high-temperature limit $\zeta/\beta \rightarrow \infty$ is more subtle, and the quantization along the crossed channel sheds light on the underlying physics. Let us first analyze the behavior of Eq. \eqref{eq:corr_sigma_zeta} in the limit above
\be\label{eq:corr_sigma_zeta_high}
\la \sigma_b(0)\tilde{\sigma}(\zeta,\bar{\zeta})\ra \simeq \l\frac{\pi}{\beta}\r^{\Delta+\Delta_b} 2^{2\Delta} \exp\l -2\pi \Delta\frac{\zeta}{\beta}\r.
\ee
As long as $\Delta \neq 0$, the two-point function decays exponentially fast to zero with a rate independent on $\Delta_b$. We interpret the last term in Eq. \eqref{eq:corr_sigma_zeta_high} as a bulk contribution, originating from the correlation between the degrees of freedom in $[0,\zeta]$ and the thermal bath.
In contrast, we dubbed the prefactor $\sim \frac{1}{\beta^{\Delta+\Delta_b}}$, which does not scale with $\zeta$, as a boundary term. Indeed, the latter is expected to be related to the short-range correlations between degrees of freedom across the entangling points $0,\zeta$ respectively, localized around a region of length $\beta$.

To better understand the origin of these bulk/boundary contributions in Eq. \eqref{eq:corr_sigma_zeta_high}, it is particularly convenient to quantize the half-cylinder along the spatial direction (crossed channel). In this way, we equivalently describe our physical system via a CFT for a system of finite size $\beta$. We denote the corresponding Hamiltonian by\footnote{The Casimir energy, which contributes as a constant shift $-\frac{c}{12}\frac{2\pi}{\beta}$ to the Hamiltonian, does not play any role in our discussion, and it is discarded explicitly. Indeed, it would simplify for the thermal expectation values we consider below, whenever the thermal states are properly normalized.
}
\be\label{eq:Hprime}
H' = \frac{2\pi}{\beta}\l L_0 + \bar{L}_{0}\r.
\ee
The ground state of $H'$ with periodic boundary condition is denoted by $\ket{0}_\beta$, and it is propagated until $\tilde{\sigma}(\zeta,\bar{\zeta})$ is inserted. Then, since the branch-cut introduced by the twist fields induced a change of boundary conditions, the propagation from $\tilde{\sigma}(\zeta,\bar{\zeta})$ to $\sigma_b(0)$ occurs via the Hamiltonian $H'_{\text{twist.}}$  with twisted boundary conditions \footnote{Both $H'$ and $H'_{\text{twist.}}$ are written in terms of the Virasoro modes as \eqref{eq:Hprime}. The difference is given by the Virasoro representation they act on.}. Equivalently, the defect line introduced by  $\tilde{\sigma}(\zeta,\bar{ \zeta})$ selects the states from the relevant twisted sector.

\begin{figure}[t]
    \centering
\includegraphics[width=0.7\linewidth]{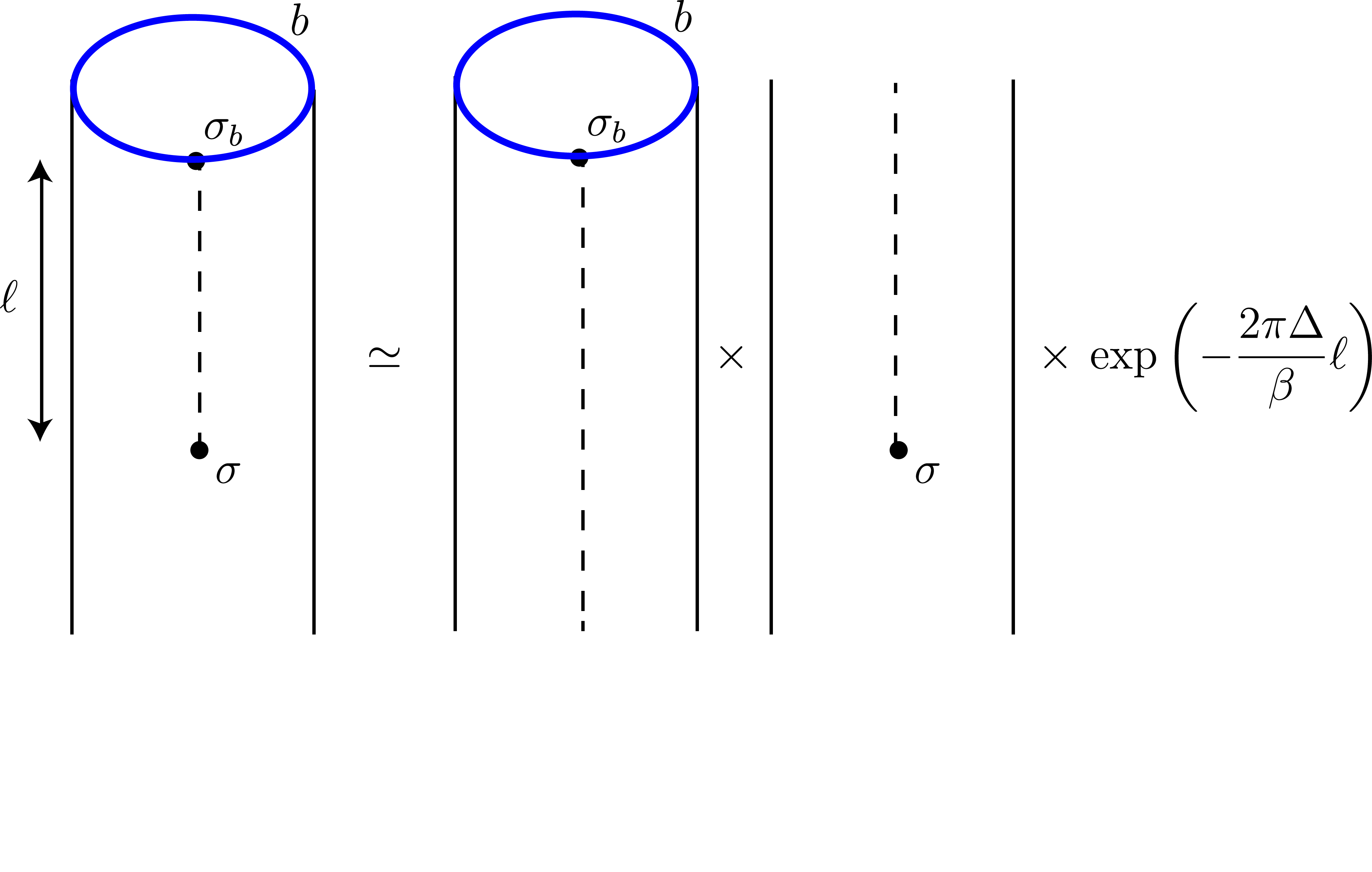}
    \caption{High-temperature limit of the correlation function \eqref{eq:corr_sigma_zeta}. For large $\ell$, an exponential decay is observed, which is related to the energy difference between the untwisted and twisted ground state. The blue line, with the additional insertion of $\sigma_b$, corresponds to the boundary state ${}_\beta\bra{b}\sigma_b(0)$. }
    \label{fig:Semi_High}
\end{figure}

Finally, the propagation ends up with a boundary state denoted by
\be\label{eq:BState_sigma}
{}_{\beta}\bra{b}\sigma_b(0).
\ee
The latter is obtained starting from the usual boundary state $\bra{b}$, corresponding to $b$, and keeping the leading term arising from the insertion of $\sigma(0)$ at the boundary. We put everything together, and we express
\be\label{eq:corr_cr_channel}
\la \sigma_b(0)\tilde{\sigma}(\zeta,\bar{\zeta})\ra = {}_{\beta}\bra{b}\sigma_b(0)\tilde{\sigma}(\zeta,\bar{\zeta})\ket{0}_\beta = {}_{\beta}\bra{b}\sigma_b(0) e^{-\zeta H'_{\text{twist.}}} \tilde{\sigma}(0)\ket{0}_\beta,
\ee
where the normalization of the boundary state $_\beta\la b|0\ra_\beta=1$ is chosen, and we have passed to the Schrodinger picture for the twist operators in the last equality in \eqref{eq:corr_cr_channel}.
At this point, we expand $H'_{\text{twist.}}$ on the basis of its eigenstates in Eq. \eqref{eq:corr_cr_channel}. This approach is predictive since the spectrum of $H'_{\text{twist.}}$ is discrete with a level spacing of order $\sim \beta^{-1}$, and for large $\zeta/\beta$ only the low-energy spectrum contributes to Eq. \eqref{eq:corr_cr_channel}. Assuming explicitly that $\sigma$ is a primary twist field, we have that the leading non-vanishing term in Eq. \eqref{eq:corr_cr_channel} comes from the ground state of $H'_{\text{twist.}}$, denoted by $\ket{0_\text{twist.}}_\beta$, with energy $E_{\text{twist.}}$. Therefore, Eq. \eqref{eq:corr_cr_channel} gives
\be\label{eq:corr_sigma_zeta_high_2}
\la \sigma_b(0)\tilde{\sigma}(\zeta,\bar{\zeta})\ra \simeq   {}_\beta\bra{b}\sigma_b(0)\ket{0_\text{twist.}}_\beta \times {}_\beta\bra{0_\text{twist.}}\tilde{\sigma}(0)\ket{0}_\beta \times \exp\l -\zeta E_{\text{twist.}} \r
\ee
in the large $\zeta/\beta$ limit, which has to be compared to Eq. \eqref{eq:corr_sigma_zeta_high}. We first find that the rate of exponential decay is ruled by the energy of the twisted ground state, and from Eq. \eqref{eq:corr_sigma_zeta_high_2} we get directly
\be
E_{\text{twist.}} = \frac{2\pi \Delta}{\beta}.
\ee
Moreover, we observe that the one-point functions appearing in Eq. \eqref{eq:corr_sigma_zeta_high_2}, referring to a finite-size geometry of length $\beta$, can be fixed by scale invariance, say
\be
{}_\beta\bra{b}\sigma_b(0)\ket{0_\text{twist.}}_\beta \sim \frac{1}{\beta^{\Delta_b}}, \quad {}_\beta\bra{0_\text{twist.}}\tilde{\sigma}(0)\ket{0}_\beta\sim \frac{1}{\beta^\Delta},
\ee
up to dimensionless prefactors. This finding is compatible with our previous result \eqref{eq:corr_sigma_zeta_high}. We give a pictorial representation of the construction above in Fig. \ref{fig:Semi_High}.

We end this subsection with some closing remarks. Notably, the derivation of \eqref{eq:corr_sigma_zeta} relies on conformal invariance only, and the same result, discussed here for twist fields, holds for untwisted primary fields too. In that case, $\tilde{\sigma}(\zeta,\bar{\zeta})$ interpolates between $\ket{0}_\beta$ and the excited state $\ket{\sigma}_\beta$ of the untwisted Hamiltonian $H'$ associated with $\sigma$ via state-operator correspondence.

Finally, it is worth analyzing formally the large distance behavior of Eq. \eqref{eq:corr_sigma_zeta} when $\Delta =0$, which is particularly important for the logarithmic negativity (as discussed in Sec. \ref{sec:log_neg})
\be
\la \sigma_b(0)\tilde{\sigma}(\zeta,\bar{\zeta})\ra = \l\frac{\pi}{\beta} \frac{1}{\tanh \l \frac{\pi \zeta}{\beta}\r}\r^{\Delta_b}.
\ee
An exponentially fast convergence to the saturation value $\l \frac{\pi}{\beta}\r^{\Delta_b}$ is observed, and for large $\zeta/\beta$ we expand
\be
\la \sigma_b(0)\tilde{\sigma}(\zeta,\bar{\zeta})\ra \simeq \l\frac{\pi}{\beta} \r^{\Delta_b}\l 1+2\Delta_b e^{-\frac{2\pi \zeta}{\beta}}+ \dots \r.
\ee
This simple analysis suggests that, beside the ground state $\ket{0}$, an excited state of energy $\frac{2\pi}{\beta}$ (corresponding to a field of scaling dimension $1$) arises from $e^{-\zeta H'_{\text{twist.}}}$ in Eq. \eqref{eq:corr_cr_channel} and interpolates between $\tilde{\sigma}(\zeta,\bar{\zeta})$ and $\sigma_b(0)$. These considerations are only formal (as they come from an interpretation of the exact results) as a comprehensive characterization of the boundary operator spectrum of junction CFTs is beyond the scope of this work.

\section{CFT approach: thermal effects at finite size}\label{sec:CFT_finite_size}
In this section, we explain how the approach of Sec. \ref{sec:CFT_half_line} can be adapted to deal with the finite size effects. This allows us to make a connection with the predictions of \cite{ss-08}, where a conformal junction at finite size and zero temperature was considered. It also helps to understand the regimes where the presence of an additional boundary is relevant.

We consider a one-dimensional system in a box of length $L$. Conformal boundary conditions of types $b$ and $b'$ are chosen at the two boundaries,  and we do not make any specific assumptions about $b'$. This setup is relevant for studying multiple \textit{finite} wires of length $L$ joined together at their extremal points via two defects. We study the finite temperature setup, which is described by a BCFT defined on a cylinder of height $L$, and circumference $\beta$ parametrized as
\be
\text{Re}\l \zeta\r \in [0,L), \quad \text{Im}(\zeta) \in [0,\beta).
\ee
We then attempt to study a two-point function of twist fields, to generalize the results in Eq, \eqref{eq:corr_sigma_zeta}. In particular, we consider
\be\label{eq:corr_fin_size}
\la \sigma_b(0)\tilde{\sigma}(\zeta,\bar{\zeta})\ra
\ee
as a function of $\beta,L,\zeta$ and the boundary conditions $b,b'$. Unfortunately, we are not able to get a closed formula in this case. However, salient features can be discussed in specific limits.

We first discuss the limit $\beta\gg L$, which is the low-temperature regime. Here, we study the thermal expectation value as a weighted sum over the eigenstates of the finite-size Hamiltonian\footnote{Up to the Casimir energy $-\frac{c}{24}\frac{\pi}{L}$, which does not play any role here.}
\be\label{eq:Hbb1_chiral}
H_{bb'}=\frac{\pi}{L}L_0,
\ee
which has the spectrum of a chiral CFT on a ring of length $2L$ (see Ref. \cite{Cardy-04} for additional details). We write the expectation value as
\be\label{eq:finiteLT}
\la \sigma_b(0)\tilde{\sigma}(\zeta,\bar{\zeta})\ra = \frac{\text{Tr}\l e^{-\beta H_{bb'}} \sigma_b(0)\tilde{\sigma}(\zeta,\bar{\zeta})\r}{\text{Tr}\l e^{-\beta H_{bb'}}\r}= \frac{ _{bb'}\bra{0}\sigma_b(0)\tilde{\sigma}(\zeta,\bar{\zeta})\ket{0}_{bb'}+\sum_\psi e^{-\beta E_\psi}\bra{\psi}\sigma_b(0)\tilde{\sigma}(\zeta,\bar{\zeta})\ket{\psi}}{1+\sum_\psi e^{-\beta E_\psi}},
\ee
with $\ket{0}_{bb'}$ the ground state of \eqref{eq:Hbb1_chiral} and $\psi$ its generic excited state. Here, $E_{\psi}$ represents the energy difference of $\ket{\psi}$ and $\ket{0}_{bb'}$, and it is proportional to $1/L$. While the ground state is not guaranteed to be unique, and properly fine-tuned boundary conditions can induce exact degeneracies in principle, we assume it is the case so that the $\beta \rightarrow \infty$ of the thermal state is pure.

From Eq. \eqref{eq:finiteLT} it is evident that the first contribution is the zero temperature correlation function. Assuming a non-degenerate ground state, the first correction goes like $\sim e^{-\frac{\pi\beta}{L}\Delta_\psi}$, with $\Delta_\psi$ the dimension of the chiral field associated with the lightest excited state $\ket{\psi}$. A direct calculation of $_{bb'}\bra{0}\sigma_b(0)\tilde{\sigma}(\zeta,\bar{\zeta})\ket{0}_{bb'}$, and $\bra{\psi}\sigma_b(0)\tilde{\sigma}(\zeta,\bar{\zeta})\ket{\psi}$ is nevertheless difficult. The reason is that the quantities above are, by conformal transformations, related to four-point functions in the upper half-plane, where $\sigma,\sigma_b$ and two boundary-changing operators are inserted. By Cardy's doubling trick\cite{Cardy-84}, this correlator is constructed with the conformal blocks of a $5$-point function on the complex plane, which are objects on which we do not have good analytic control.

In the high-temperature regime $\beta\ll L,\zeta$, one can employ the techniques of Sec. \ref{sec:halfline_highT}. We quantize in the crossed channel and write the correlation function as an overlap between boundary states
\be
\la \sigma_b(0)\tilde{\sigma}(\zeta,\bar{\zeta})\ra  = \frac{_{\beta}\bra{b'}e^{-(L-\zeta) H'}\tilde{\sigma}(0)e^{-\zeta H'_{\text{twist.}}}\sigma_b(0)\ket{b}_\beta}{_{\beta}\bra{b'}e^{-LH'}\ket{b}_\beta},
\ee
that is the analogue of Eq. \eqref{eq:corr_cr_channel}. We expand over the eigenstates of $H'$ and $H'_{\text{twist}.}$ and the leading term is
\be\label{eq:LeadingHightT}
\la \sigma_b(0)\tilde{\sigma}(\zeta,\bar{\zeta})\ra \simeq _{\beta}\bra{0}\tilde{\sigma}(0)\ket{0_{\text{twist.}}}_{\beta} \times _{\beta}\bra{0_{\text{twist}.}}\sigma_b(0)\ket{b}_{\beta}  \times \exp\l -\zeta E_{\text{twist.}}\r \propto \frac{1}{\beta^{\Delta+\Delta_b}} \exp\l -\frac{2\pi \Delta}{\beta}\zeta\r,
\ee
which is the same as the result \eqref{eq:corr_cr_channel}. In particular, the dependence on $b'$ is lost. This is physically compatible with the presence of a finite (small) correlation length, and one expects that a probe far from the boundary should not be affected by it.

The first non-trivial correction comes from the first excited state of $H'$, denoted by $\ket{\psi}_\beta$ with energy $\frac{2\pi \Delta_\psi}{\beta}$. Indeed, by expanding $e^{-(L-\zeta)H'}$ and keeping the contribution from $\ket{\psi}_\beta$ and $\ket{0}_\beta$ only, we get a correction to \eqref{eq:LeadingHightT} 
\begin{align}
&_{\beta}\la b' |\psi\ra_{\beta} \times_{\beta}\bra{\psi}\tilde{\sigma}(0)\ket{0_{\text{twist.}}}_{\beta} \times _{\beta}\bra{0_{\text{twist}.}}\sigma_b(0)\ket{b}_{\beta}  \times \exp\l -\zeta E_{\text{twist.}}\r \times  \exp\l -\frac{2\pi \Delta_\psi}{\beta}(L-\zeta)\r\propto \\
&\frac{1}{\beta^{\Delta+\Delta_b}} \exp\l -\frac{2\pi \Delta}{\beta}\zeta\r \times \exp\l -\frac{2\pi \Delta_\psi}{\beta}(L-\zeta)\r,
\end{align}
where dimensionless proportionality constants, related to structure constants of the fields, have been discarded. Clearly, if $\ket{\psi}_\beta$ is degenerate, one has to sum over the states with the same energy of $\ket{\psi}_\beta$. In general, our argument shows that, in the regime considered above, the finite size corrections appear as a power series of $\exp\l -\frac{2\pi(L-\zeta)}{\beta}\r$.

The previous analysis refers to an insertion of the $\tilde{\sigma}$ in a generic bulk point $\zeta \in (0,L)$. A slightly different mechanism occurs if the field above is inserted at the boundary point $\zeta = L$, and we discuss it below.

\subsection{Correlations between boundary twist fields}

When we set $\zeta = L$, the operator $\tilde{\sigma}$ is replaced by a boundary twist field $\tilde{\sigma}_{b'}(L)$ with scaling dimension $\Delta_{b'}$. Therefore, the two-point function is expressed as
\be
\la \sigma_b(0)\tilde{\sigma}_{b'}(L)\ra,
\ee
in analogy with Eq. \eqref{eq:corr_fin_size}.

Let us first consider the small temperature limit $\beta\gg L$. We expand over the eigenstates of the direct channel, as in Eq. \eqref{eq:finiteLT}, and we get
\be\label{eq:ins_bound_low_T}
\la \sigma_b(0)\tilde{\sigma}_{b'}(L)\ra = \frac{ _{bb'}\bra{0}\sigma_b(0)\tilde{\sigma}_{b'}(L)\ket{0}_{bb'}+\sum_\psi e^{-\beta E_\psi}\bra{\psi}\sigma_b(0)\tilde{\sigma}_{b'}(L)\ket{\psi}}{1+\sum_\psi e^{-\beta E_\psi}}.
\ee
From scaling arguments, the matrix elements appearing in the expression above depend on the length $L$ as follows
\be
\bra{\psi}\sigma_b(0)\tilde{\sigma}_{b'}(L)\ket{\psi} \propto \ _{bb'}\bra{0}\sigma_b(0)\tilde{\sigma}_{b'}(L)\ket{0}_{bb'} = \frac{1}{L^{\Delta_b+\Delta_{b'}}},
\ee
where universal proportionality constants have been discarded \footnote{The latter can be related to $4$-point functions of boundary operators in the upper half-plane. However, we do not attempt to calculate them here.}. Therefore, we organize the term in Eq. \eqref{eq:ins_bound_low_T} as follows
\be
\la \sigma_b(0)\tilde{\sigma}_{b'}(L)\ra = \frac{1}{L^{\Delta_b+\Delta_{b'}}}\l 1+ O(e^{-\pi \Delta_\psi\frac{\beta}{L}})\r,
\ee
where a series in $e^{-\pi\frac{\beta}{L}}$ is present inside the parenthesis and $\Delta_\psi$ is the dimension of the lightest excited state. At strictly zero temperature and purely reflective boundary conditions $b'$ at $\zeta=L$, one recovers the physical features discussed in Refs. \cite{ss-08,gm-17,cmc-22,cmc-22a,bcp-23}, and the algebraic decay of the correlation function within $L$ corresponds to the ubiquitous logarithmic growth of entanglement entropy (see Sec. \ref{sec:Ent_measures} for details). We also discover that setting permeable boundary conditions along $b'$ leads to an additive contribution to entropy, at least at the leading order. Demonstrating this assertion at the microscopic level is challenging; for instance, the technique presented in Ref. \cite{cmv-12} cannot be directly adapted. However, this remarkable outcome is a direct consequence of field theory, and a thorough investigation of its details is deferred to potential future studies.

In the regime of high-temperature $\beta\ll L$, we employ the quantization along the crossed channel as in 
 Eq. \eqref{eq:corr_cr_channel}
\begin{align}
\nonumber
&\la \sigma_b(0)\tilde{\sigma}_{b'}(L)\ra  = \frac{_{\beta}\bra{b'}\tilde{\sigma}_{b'}(0)e^{-LH'_{\text{twist.}}}\sigma_b(0)\ket{b}_\beta}{_{\beta}\bra{b'}e^{-LH'}\ket{b}_\beta} \simeq\\
&_{\beta}\bra{b'}\tilde{\sigma}_{b'}(0)\ket{0_{\text{twist.}}}_{\beta} \times _{\beta}\bra{0_{\text{twist}.}}\sigma_b(0)\ket{b}_{\beta}  \times \exp\l -L E_{\text{twist.}}\r \propto \frac{1}{\beta^{\Delta_{b'}+\Delta_b}} \exp\l -\frac{2\pi \Delta}{\beta}L\r.
\end{align}
While the above equation is reminescent of  \eqref{eq:LeadingHightT}, and the exponential decay of correlations is equivalent (up to $\zeta \rightarrow L$), there is now an extra dependence on the boundary dimension $\Delta_{b'}$. This is the major novelty, which comes precisely from the insertion of the twist field at the boundary $b'$.

\section{Entanglement measures}\label{sec:Ent_measures}

In this section, we discuss the relation between the thermal correlation functions of twist fields, characterized in Sec. \ref{sec:CFT_half_line}, and the entanglement between spatial regions. We focus on the R\'enyi entropies \cite{cc-04} and logarithmic negativity \cite{Plenio-05} of a pair of intervals attached to a defect. We aim to understand the role of the boundary conditions at the defect, and how they affect the entanglement at finite temperature.

For simplicity, let us consider a conformal junction made of two semi-infinite wires connected together at a single point. This system is described by a (boundary) CFT denoted by \cite{gm-17}
\be
\text{CFT} = \text{CFT}_1 \otimes \text{CFT}_2.
\ee
We parametrize the points of the junction via a pair\cite{cmc-22}
\be
(x,j), \quad x \in [0,+\infty), \quad j=1,2,
\ee
with $j$ labelling the two wires and $x$ their spatial points. We impose conformal invariant boundary conditions $b$ at $x=0$, which, in general, couple the two wires.

We replicate the theory $n$ times, that is the starting point to compute the entanglement measures mentioned above via replica trick, and we consider the theory
\be
\text{CFT}^{\otimes n} = \l\text{CFT}_1 \otimes \text{CFT}_2\r^{\otimes n}.
\ee
We introduce an additional label
\be
k = 1,\dots,n,
\ee
to parametrize the $n$ replicas.
A family of twist fields in $\text{CFT}^{\otimes n}$ can be defined\footnote{Strictly speaking, the twist fields are not local fields of the replica theory $\text{CFT}^{\otimes n}$. Still, an orbifold model $\text{CFT}^{\otimes n}/\mathbb{Z}_n$ can be constructed, promoting the replica permutation to global symmetry, and the twist fields would appear as local fields of the orbifold, as explained in detail in Ref. \cite{eir-22a}.}  such that their insertion gives introduce branch cuts on each of the wires, implementing the appropriate replica shifts. For instance, following closely Ref. \cite{ce-23}, we require that the twist fields\footnote{We have opted to use a different notation for the twist fields of the junction orbifold CFT to stress that the discussion of this section and Sec. \ref{sec:Numerics} is a particular case of the general results derived previously.} $\mathcal{T}^j,\tilde{\mathcal{T}}^j$ are the lightest fields satisfying
\be\label{eq:Twist_fields}
\begin{split}
\mathcal{T}^{j}(x) \mathcal{O}_{j',k}(x') = \begin{cases} \mathcal{O}_{j',k+1}(x')\mathcal{T}^{j}(x), \quad x<x',\quad j=j'\\ \mathcal{O}_{j',k}(x')\mathcal{T}^{j}(x), \quad \text{otherwise}, \end{cases} \\
\tilde{\mathcal{T}}^{j}(x) \mathcal{O}_{j',k}(x') = \begin{cases} \mathcal{O}_{j',k-1}(x')\tilde{\mathcal{T}}^{j}(x), \quad x<x',\quad j=j'\\ \mathcal{O}_{j',k}(x')\tilde{\mathcal{T}}^{j}(x), \quad \text{otherwise}, \end{cases}
\end{split}
\ee
with $\mathcal{O}_{j,k}(x)$ being a generic local operator inserted at position $x$ in the $k$th replica of the $j$th wire.

These fields are the building blocks to construct entanglement measures, as we show explicitly below.
For the sake of clarity, we focus on the regions
\be
A = \{(x,j) | \  x \in[0,\ell] , j=1\}, \quad B = \{(x,j) |\  x \in[0,\ell] , j=2\},
\ee
being segments of length $\ell$ attached to the defect point, and we probe their quantum correlations at finite temperature.

\subsection{R\'enyi entropy}\label{sec:Renyi_entropy}

Here, we compute the R\'enyi entropies of $A,B$ and their union $A\cup B$. Given $\rho_A$ the reduced density matrix (RDM) of $A$, we express the $n$th R\'enyi entropy of $A$ as\cite{cc-04,ce-23}
\be
S_n(A) = \frac{1}{1-n}\log \text{Tr}\l \rho^n_A\r = \frac{1}{1-n}\log \l\varepsilon^{\Delta + \Delta_b}\la \mathcal{T}^1(0)\tilde{\mathcal{T}}^1(\ell)\ra\r,
\ee
where $\Delta$ and $\Delta_b$ are the bulk/boundary scaling dimensions of $\mathcal{T}^1$, $\varepsilon$ plays the role of an arbitrary ultraviolet cutoff and we have dropped any non-universal constants. We remark that $\mathcal{T}^1(0)$ is a boundary field, and its scaling properties depend on the choice boundary conditions $b$ explicitly: for convenience, we omit the index $b$ employed in Sec. \ref{sec:CFT_half_line}. We also refer to $\la \dots\ra$ as the thermal state of the half-line at temperature $\beta^{-1}$, that is the state we aim to probe.
At this point, we employ directly the CFT result obtained in Sec. \ref{sec:CFT_half_line}, and, from Eq. \eqref{eq:corr_sigma_zeta}, we get
\be\label{eq:Sn_prediction}
S_n(A) = \frac{\Delta+\Delta_b}{n-1}\log \frac{\beta}{\varepsilon} +\frac{1}{n-1}\log \l \cosh\l \frac{\pi \ell}{\beta}\r^{\Delta-\Delta_b} \sinh\l \frac{\pi \ell}{\beta}\r^{\Delta+\Delta_b}\r + \text{const.}.
\ee
The bulk scaling dimension $\Delta$ depends only on the central charge $c_1$ of $\text{CFT}_1$, and its value is known \cite{cc-09}
\be\label{eq:Delta}
\Delta = \frac{c_1}{12}\l n-\frac{1}{n}\r.
\ee
On the other hand, the boundary dimension $\Delta_b$, which characterizes the quantum fluctuation across the defect, depends non-trivially on the theory and the boundary conditions $b$: its explicit value is related to the logarithmic divergence of the ground state entropy with defect \cite{ce-23}, that is known for free bosons and fermions (see e.g. Refs. \cite{ss-08,bb-15}). Since we want to keep the discussion general, we refrain from characterizing it further.

In the high-temperature limit, say for a large interval $\ell/\beta\gg 1 $, we get from \eqref{eq:Sn_prediction}
\be
S_n(A) \simeq \frac{\Delta+\Delta_b}{n-1}\log \frac{\beta}{\varepsilon} + \frac{2\pi \Delta}{n-1}\frac{\ell}{\beta}+\text{const.}
\ee
We interpret the first contribution as an area term arising from the fluctuation across the entangling points $(x,j)=(0,1)$ and $(\ell,1)$ respectively. The second contribution is instead a volume term, whose origin is traced back to the correlations between the system and the thermal bath.

To isolate the entanglement between the regions $A$ and $B$, and discard their correlations with the fictitious bath and the complement $\overline{A\cup B}$, one can compute their R\'enyi mutual information, defined as
\be\label{eq:mut_info}
I_n(A,B) \equiv S_n(A)+S_n(B)-S_n(A\cup B),
\ee
which gives the mutual information \cite{wvhc-08} in the replica limit $n\rightarrow 1$.

The computation of $S_n(B)$ is analogous to $S_n(A)$ up to the substitution $\mathcal{T}^1 \rightarrow \mathcal{T}^2$. In particular, the bulk scaling dimension of $\mathcal{T}^2$ is
\be
\frac{c_2}{12}\l n-\frac{1}{n}\r,
\ee
while its boundary scaling dimension $\Delta_b$ is the same as the one of $\mathcal{T}^1$. One easy way to check the validity of this statement is the following. The ground state entropy shared by the two wires of finite length $L$, can be equivalently expressed by tracing out the first or the second wire. In particular, one can compute the latter as an expectation value of $\mathcal{T}^1(0)\tilde{\mathcal{T}}^1(L)$ or $\mathcal{T}^2(0)\tilde{\mathcal{T}}^2(L)$ in the zero temperature state. If one chooses purely reflective boundary condition at $x=L$, the dimension of $\tilde{\mathcal{T}}^j(L)$ is zero, and we get the equality of the dimensions of $\mathcal{T}^1(0)$ and $\mathcal{T}^2(0)$.

To express $S_n(A\cup B)$, we first consider the twist field
\be
\mathcal{T}(x) = (\mathcal{T}^1\mathcal{T}^2)(x),
\ee
obtained from the fusion of $\mathcal{T}^1$ and $\mathcal{T}^2$. $\mathcal{T}$ acts as a replica shift $k \rightarrow k+1$ on the half-line $(x,\infty)$ for both the wires, and it can be considered as the 'standard twist field' associated with the replica model $\text{CFT}^{\otimes n}$. Its bulk scaling dimension, that we denote here by $\Delta'$, is therefore
\be
\Delta' = \frac{c_1+c_2}{12}\l n-\frac{1}{n}\r,
\ee
with $c_1+c_2$ the central charge of $\text{CFT} = \text{CFT}_1\otimes \text{CFT}_2$, while its boundary scaling dimension vanishes identically \footnote{In Ref. \cite{cc-09}, and forthcoming literature, the insertion of the corresponding boundary twist field is even discarded, and it is formally expressed as the identity operator.}. In particular, since $\Delta'$ matches the sum of the dimensions of $\mathcal{T}^1$ and $\mathcal{T}^2$, no UV divergences arise by their fusion: this is not surprising, the two species of fields are decoupled in the bulk.

We express the R\'enyi entropy of $A\cup B$ as
\be\label{eq:Sn_sym_int}
\begin{split}
S_n(A\cup B) = \frac{1}{1-n}\log \text{Tr}\l \rho^n_{A\cup B}\r = \frac{1}{1-n}\log \l\varepsilon^{\Delta' }\la \mathcal{T}(0)\tilde{\mathcal{T}}(\ell)\ra\r = \\
\frac{\Delta'}{n-1}\log \frac{\beta}{\varepsilon} +\frac{\Delta'}{n-1}\log \l \cosh\l \frac{\pi \ell}{\beta}\r\sinh\l \frac{\pi \ell}{\beta}\r\r + \text{const.}
\end{split}.
\ee
As expected, the result of \eqref{eq:Sn_sym_int} is compatible with the prediction in \cite{Mintchev:2020jhc} for the same bipartition, in the limit $\beta\gg \ell$ - up to an overall constant which we did not evaluate in the present work.

Putting everything together in the definition \eqref{eq:mut_info}, we express the R\'enyi mutual information between $A$ and $B$ as
\be\label{eq:MI_pred}
I_n(A,B) = \frac{2\Delta_b}{1-n}\log \l\frac{\pi \varepsilon}{\beta} \frac{1}{\tanh\l \frac{\pi \ell}{\beta}\r}\r.
\ee
As physically expected, in the large $\ell$ limit $I_n(A,B)$ converges to a finite value, which depends on the boundary conditions at the entangling point $x=0$ as
\be
\underset{\ell \rightarrow \infty}{\lim} I_n(A,B) = \frac{2\Delta_b}{n-1}\log \frac{\beta}{\pi \varepsilon}.
\ee
In particular, both the extensive terms of the entropy, coming from the bath, and the area term associated with $x=\ell$, are canceled in the combination \eqref{eq:mut_info} defining the mutual information. For this reason, the mutual information \eqref{eq:MI_pred} captures correctly and isolates the correlations across the defect.

\subsection{Logarithmic negativity}\label{sec:log_neg}

In this section, we compute the logarithmic negativity between $A$ and $B$ via replica trick, following Ref \cite{cct-12,cct-14}. To do so, we first introduce the R\'enyi negativity as
\be
\mathcal{E}_{n_e} = \log \text{Tr}\l \l\rho^{T_B}_{A\cup B}\r^{n_e}\r,
\ee
with $\rho^{T_B}_{A\cup B}$ the partial transposition (wrt $B$) of the RDM of $A\cup B$, and $n_e$ an even integer. Then, we perform the analytic continuation $n_e\rightarrow 1$ to get the logarithmic negativity $\mathcal{E}$ as
\be
\mathcal{E} \equiv \log \text{Tr}\l |\rho^{T_B}_{A\cup B}|\r = \underset{n_e\rightarrow 1}{\lim} \mathcal{E}_{n_e}.
\ee

Whenever $n_e$ is even, one can compute $\mathcal{E}_{n_e}$ as follows. One starts from the replica model $\text{CFT}^{\otimes n_e}$, and then insert a branch-cut implementing the replica shift $k\rightarrow k+1$ along $A$ and the shift $k\rightarrow k-1$ along $B$. We do so via the twist field
\be
\sigma(x)\equiv (\mathcal{T}^1 \tilde{\mathcal{T}}^2)(x),
\ee
and we express
\be
\mathcal{E}_{n_e} = \log \l\varepsilon^{\hat{\Delta}+\hat{\Delta}_b} \la \sigma_b(0)\tilde{\sigma}(\ell)\ra\r,
\ee
where $\hat{\Delta},\hat{\Delta}_b$ are the bulk/boundary scaling dimensions of $\sigma$. Since the two wires are decoupled in the bulk, $\hat{\Delta}$ is simply given by
\be\label{eq:bulk_dim_neg}
\hat{\Delta} = \frac{c_1+c_2}{12}\l n-\frac{1}{n}\r,
\ee
that is the sum of the dimension of $\mathcal{T}^1$ and $\tilde{\mathcal{T}}^2$.
$\hat{\Delta}_b$ is non-trivial and depends on both the boundary conditions $b$ and $n_e$. However, for the specific case of a two-wire geometry under analysis, one can show that
\be\label{eq:neg_MI_rel}
\hat{\Delta}_b|_{n_e} = 2\Delta_b|_{n=n_e/2},
\ee
with $\Delta_b$ the boundary dimension of $\mathcal{T}_1$ (or $\mathcal{T}_2$) with $n$ replicas. Eq. \eqref{eq:neg_MI_rel} follows from a general equivalence between the logarithmic negativity and the $1/2$-R\'enyi entropy \cite{cct-12} for bipartite systems (see also Ref. \cite{ge-20} for a related discussion on the ground state negativity).

Using the CFT prediction \eqref{eq:corr_sigma_zeta}, and the property $\underset{n_e\rightarrow 1}{\lim} \hat{\Delta} =0$ (see Eq. \eqref{eq:bulk_dim_neg}), one finally obtains the logarithmic negativity between $A$ and $B$
\be\label{eq:Anal_E}
\mathcal{E} = \l\underset{n_e\rightarrow 1}{\lim} \hat{\Delta}_b\r\times \log \l\frac{\pi \varepsilon}{\beta} \frac{1}{\tanh\l \frac{\pi \ell}{\beta}\r}\r.
\ee
From its explicit expression, one immediately realizes that the dependence of $\mathcal{E}$ on $\ell$ and $\beta$ is the same as $I_n(A,B)$ in Eq. \eqref{eq:MI_pred}, and the two entanglement measures differ by a proportionality constant only. Specifically, from the relation \eqref{eq:neg_MI_rel}, we get
\be\label{eq:eq:neg_MI_rel_1}
I_{1/2}(A,B) = 2\mathcal{E},
\ee
with $I_{1/2}(A,B)$ in Eq. \eqref{eq:MI_pred}. The relation above has occurred already in the literature in the context of quenches \cite{ac-19}. While \eqref{eq:eq:neg_MI_rel_1} can be proven with elementary methods\footnote{From the Schmidt decomposition of a pure state $\rho_{A\cup B}$ one can show that $\text{Tr}\l(\rho^{T_B}_{A\cup B})^{n_e}\r = \text{Tr}\l(\rho_{A\cup B})^{n_e/2}\r^2$, as explained in \cite{cct-12}.} whenever $\rho_{A\cup B}$ is a pure state, it is not always valid, and simple counterexamples can be provided. In particular, it is not clearly a priori why \eqref{eq:eq:neg_MI_rel_1} should hold exactly in our context, as the state we are considering is not pure. However, we believe that its validity strongly relies on the two-wire geometry considered above. For the sake of completeness, we mention that already for a three-wire geometry at zero temperature, analysed in detail in \cite{cmc-22,cmc-22a}, the mutual information between two wires does not obey \eqref{eq:eq:neg_MI_rel_1}.
To conclude this section, we note that the above discussion can be generalized straightforwardly to $M>2$ wires. While the approach of Sec. \ref{sec:CFT_half_line} can be easily adapted, additional replica twist fields, acting non-trivially on subsets of wires (as in Eq. \eqref{eq:Twist_fields}), should be considered.

\section{Lattice and numerics}\label{sec:Numerics}

We consider a lattice junction of free fermions, made by $M$ wires coupled at a single point via a conformal defect, that is a generalization of the model in Ref. \cite{ep-12,pe-12} where two wires were studied. We parametrize the point of the junctions via a semi-integer spatial index
\be
x = \frac{1}{2},\dots,L-\frac{1}{2},
\ee
corresponding to the position along on each wire, and an index
\be
j=1,\dots,M
\ee
labeling the wires. The Hamiltonian of the system is
\be\label{eq:H_lattice}
H = -\frac{1}{2}\sum^{M}_{j=1}\sum^{L-3/2}_{x=1/2} \Psi^\dagger_j(x)\Psi_j(x+1) + \text{c.c.} -\frac{1}{2}\sum^{M}_{j,j'=1} S_{jj'}\Psi^\dagger_j(1/2)\Psi_{j'}(1/2),
\ee
where $S$ is a $M\times M$ matrix that is unitary and hermitian, while $\Psi_j(x),\Psi^\dagger_j(x)$ are annihilation creation fermionic operators satisfying the usual anticommutation relations
\be
\{\Psi_j(x),\Psi_{j'}(x')\} = 0, \quad \{\Psi_j(x),\Psi^\dagger_{j'}(x')\} = \delta_{jj'}\delta_{xx'}.
\ee
The first term in \eqref{eq:H_lattice} is the hopping, which does not couple wires, while the second one in \eqref{eq:H_lattice} couples them explicitly. It is possible to show that $S$ is precisely the scattering matrix of the problem\footnote{To do so, it is sufficient to study the single-particle scattering problem. One can do it without the assumption of $S$ unitary (while its hermiticity ensures the hermiticity of the Hamiltonian \eqref{eq:H_lattice}). In general, the scattering matrix is $k$-dependent; if we further assume $S^2=1$, the dependence on the momentum is lost, and, for this reason, we say that the defect is conformal.} and, as long as it is not diagonal, particles at the boundary point $x=1/2$ can hop between distinct wires.

We consider the thermal state of \eqref{eq:H_lattice} at temperature $\beta^{-1}$. To characterize that, we compute the two-point function denoted by
\be
C_{jj'}(x,x') \equiv \la \Psi^\dagger_{j'}(x')\Psi_j(x)\ra = \frac{ \text{Tr}\l e^{-\beta H}\Psi^\dagger_{j'}(x')\Psi_j(x)\r }{\text{Tr}\l e^{-\beta H}\r}.
\ee
We briefly sketch the procedure below
\begin{itemize}
\item We perform a unitary transformation which diagonalizes $S$ while keeping the hopping term untouched and expressing the Hamiltonian in terms of decoupled nonphysical fields (see Ref. \cite{cmc-22,cmc-22a} for details). 
\item The eigenmodes of the nonphysical fields are combinations of plane waves, where Neumann/Dirichlet boundary conditions are present at the defect point. The latter corresponds to the eigenvalues $\pm 1$ of the matrix $S$ respectively (we recall that we assume $S^2 = 1$).
\item We construct the thermal correlation functions of the nonphysical fields, and then, after a change of basis, we express the ones of the original fields $\Psi_j(x)$.
\end{itemize}
The final result is
\be\label{eq:Corr_matrix}
C_{jj'}(x,x') = \l\frac{1+S}{2}\r_{jj'} C_{N}(x,x') + \l\frac{1-S}{2}\r_{jj'} C_{D}(x,x'),
\ee
with
\be\label{eq:CN_CD}
C_{N}(x,x') = \frac{2}{L} \sum_k \cos(kx)\cos(kx')n_\beta(k), \quad C_{D}(x,x') = \frac{2}{L} \sum_k \sin(kx)\sin(kx')n_\beta(k),
\ee
and $n_\beta(k)$ the thermal occupation
\be
n_\beta(k) = \frac{1}{e^{-\beta \cos k}+1}.
\ee
The sums appearing in the definition of the Neumann/Dirichlet correlation functions in Eq. \eqref{eq:CN_CD} depend on a set of $L$ quantized momenta $k \in [0,\pi]$, fixed by the boundary terms of \eqref{eq:H_lattice} at $x=1/2$ and $x=L-1/2$, that we do not specify analytically. If we consider the infinite-volume limit
\be
L\rightarrow \infty, \quad x,x' \text{ fixed},
\ee
and we replace
\be
\frac{1}{L}\sum_k \rightarrow \int^{\pi}_0 \frac{dk}{\pi},
\ee
in Eq. \eqref{eq:CN_CD}, so the quantization of momenta does not play any role. For the sake of convenience, we define the kernel
\be\label{eq:Kbeta}
\mathcal{K}_\beta(x) \equiv \int^{\pi}_{-\pi} \frac{dk}{2\pi} n_\beta(k)e^{ikx},
\ee
and we finally express
\be
C_N(x,x') = \mathcal{K}_\beta(x-x') + \mathcal{K}_\beta(x+x'), \quad C_D(x,x') = \mathcal{K}_\beta(x-x') -\mathcal{K}_\beta(x+x')
\ee
in the large volume limit. At zero-temperature i.e. $\beta\rightarrow \infty$, the equation \eqref{eq:Kbeta} gives the sine-kernel
\be
\mathcal{K}_\infty(x)  = \frac{\sin \frac{\pi}{2}(x-x')}{\pi(x-x')},
\ee
at half-filling (see Ref. \cite{fc-11}), corresponding to the Fermi point $k_F = \pi/2$, and algebraic decay of correlations is present. While, as soon as $\beta$ is finite, the kernel $\mathcal{K}_\beta(x)$ decays exponentially to zero within a typical correlation length $\sim \beta$.

We finally observe that, if we perform the bulk limit, say
\be
x,x' \rightarrow +\infty, \quad x-x' \text{ fixed}
\ee
in Eq. \eqref{eq:Corr_matrix}, we obtain a fast convergence of the two-point function to the value
\be
C_{jj'}(x,x') \simeq \delta_{jj'}\mathcal{K}_\beta(x-x').
\ee
In this limit, the thermal state of the junction is indistinguishable from the thermal state in the infinite volume limit, and the boundary does not play any role. While the latter property matches a simple physical intuition, it is not completely obvious, since non-trivial stationary states with long-range correlations between the wires can emerge in principle from the defect (see e.g. \cite{bf-16,lsp-19,gls-23,fg-23,csrc-23}).

\subsection{R\'enyi entropies}\label{sec:Num_Renyi}

To compute the R\'enyi entropy from the correlation functions, we follow Ref. \cite{Peschel-03,Peschel-09}. We first construct the correlation matrix $C$, reported in Eq. \eqref{eq:Corr_matrix} for the half-line geometry, and we consider its restriction over a spatial subset $A$ which we denote as $C_A$. The R\'enyi entropy is thus expressed as
\be
S_n(A) = \frac{1}{1-n}\log\text{Tr}\l \l C_A\r^n +\l 1-C_A\r^n\r,
\ee
and the von Neumann entropy, obtained as the limit $n\rightarrow 1$, is
\be
S_1(A) = -\log\text{Tr}\l  C_A \log C_A +  (1-C_A) \log (1-C_A)\r.
\ee

\begin{figure}[thp]
    \centering
    \includegraphics[width=0.5\textwidth]{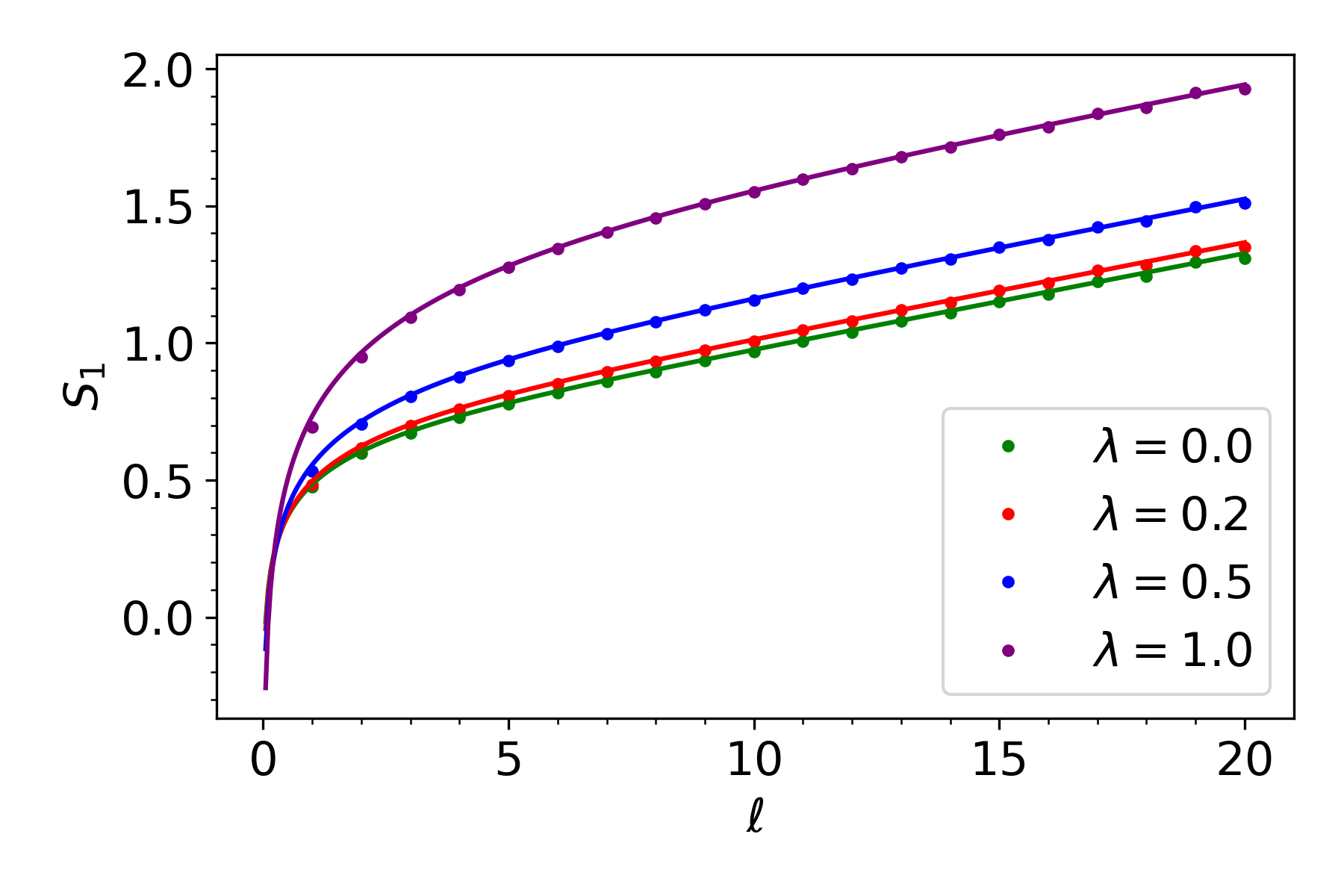}
    \caption{Entanglement entropy $S_1$ of an interval of length $\ell$ attached to the defect in a two-wire geometry. The dots are numerical data, while the analytical results (from Eq. \eqref{eq:Sn_prediction}) are continuous lines: an overall additive constant, not predicted by Eq. \eqref{eq:Sn_prediction} and corresponding to a non-universal cut-off, has been fitted. Some values of the scattering amplitude $\lambda$ have been considered ($\lambda=0,0.2,0.5,1$). A linear growth of the entropy, independent of $\lambda$, is observed at large $\ell$ and a non-trivial crossover occurs at finite $\ell/\beta$.}
    \label{fig:S1_1int}
\end{figure}

We test numerically our prediction Eq. \eqref{eq:Sn_prediction} as follows. We choose $M=2$, a junction with two wires, and we consider the following scattering matrix
\be\label{eq:Scatt_Mat}
S = \begin{pmatrix} \sqrt{1-\lambda^2} & \lambda \\ \lambda & -\sqrt{1-\lambda^2}\end{pmatrix},
\ee
parametrized by the scattering amplitude $\lambda \in [0,1]$. The system above has been studied in Refs. \cite{ep-12,ce-23,ce-23a}.
We first focus on an interval of length $\ell$ attached to the defect that belongs to the first wire: it corresponds to the sites $x \in [1/2,\dots,\ell-1/2]$ and $j=1$ in the Hamiltonian \eqref{eq:H_lattice}. To employ Eq. \eqref{eq:Sn_prediction}, we need to specify the values of scaling dimensions for the bulk/boundary twist field $\mathcal{T}^1$, defined in Eq. \eqref{eq:Twist_fields}. In particular, we are interested in the limit $n\rightarrow 1$. Since the underlying CFT is the Dirac massless theory, whose central charge is $c=1$ \cite{dms-97}, we have from Eq. \eqref{eq:Delta}
\be
\Delta = \frac{1}{12}\l n-\frac{1}{n}\r, \quad \underset{n\rightarrow 1}{\lim}\frac{1}{n-1}\Delta = \frac{1}{6}.
\ee
The boundary scaling dimension $\Delta_b$, is non-trivial and it depends explicitly on $\lambda$. Its value has been reported in \cite{ce-23}, and it is related to the so-called \textit{effective central charge} computed in \cite{pe-12}. In particular, the latter is defined as
\be
c_{\text{eff}}=-\frac{6}{\pi^2} \bigg\{ \Big[ (1+\lambda) \log (1+\lambda) + (1-\lambda) \log (1-\lambda) \Big] \log \lambda
+ (1+\lambda) \mathrm{Li_2}(-\lambda) + (1-\lambda) \mathrm{Li_2}(\lambda) \bigg\} \, ,
\ee
with $\mathrm{Li_2}(\lambda)$ the polylogarithm, and $\Delta_b$ satisfies
\be
\underset{n\rightarrow 1}{\lim}\frac{1}{n-1}\Delta_b = \frac{c_{\text{eff}}}{6}.
\ee
In Fig. \ref{fig:S1_1int} we plot the numerical data for the von Neumann entropy obtained with $L=30$ and $\beta = 30$ as a function of the subsystem size $\ell$. Remarkably, despite the thermal length not being particularly small compared to the system size, our analytical predictions reproduce faithfully the data.

\begin{figure}[thp]
    \centering
    \includegraphics[width=0.5\textwidth]{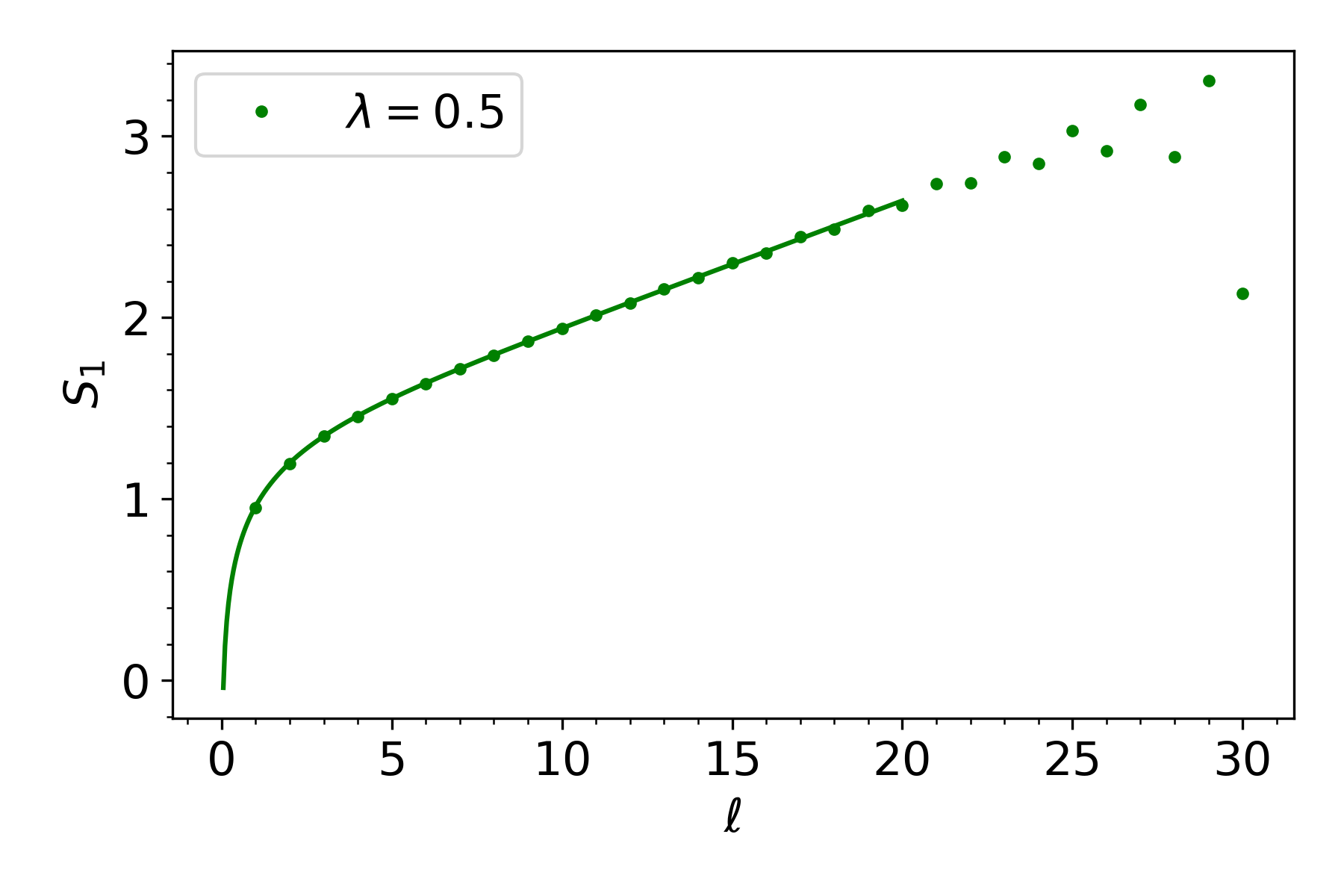}
    \caption{Entanglement entropy $S_1$ of an interval of length $2\ell$ placed symmetrically across the defect. The dots/line are numerical/analytical results respectively: an overall additive constant, not predicted by Eq. \eqref{eq:Sn_prediction} and corresponding to a non-universal cut-off, have been fitted. We choose $L=30,\beta = 30, \lambda= 0.5$ and we study the behavior of $S_1$ for $\ell \in [1,L]$.}
    \label{fig:S1_2int}
\end{figure}

We now consider an interval $A$ placed symmetrically across the defect. Specifically, $A$ contains the sites $x \in [1/2,\dots,\ell-1/2]$ in both the $j=1$ and $j=2$ wires . To represent its R\'enyi entropy in the folded picture, it is sufficient to consider the twist field $\mathcal{T} \equiv \mathcal{T}^1 \times \mathcal{T}^2$, that is, just the \textit{standard twist field} associated with the theory $\text{CFT}_1 \otimes \text{CFT}_2$. In this case,
\be
\Delta = \frac{2}{12}\l n-\frac{1}{n}\r, \quad \underset{n\rightarrow 1}{\lim}\frac{1}{n-1}\Delta = \frac{1}{3},
\ee
since $\text{CFT}_1 \otimes \text{CFT}_2$ has central charge $2$ (two species of fermions are present). The scaling dimension of  the boundary twist field vanishes (see Sec. \ref{sec:Renyi_entropy})
\be
\Delta_b =0,
\ee
as physically expected, since the defect is not supposed to contribute to the entanglement in the geometry considered above. The prediction for $S_n$ has been reported in Eq. \eqref{eq:Sn_sym_int}. We test it numerically and, for $L=30,\beta = 30, \lambda= 0.5$, we plot the results in Fig. \ref{fig:S1_2int}. As expected, our formula Eq. \eqref{eq:Sn_sym_int}, obtained in the limit $L\rightarrow \infty$, correctly describes the growth of the entropy until $\ell \simeq L$, where finite-size effects become relevant. In addition, we mention the presence of oscillations around the analytical predictions. While their characterization is beyond the scope of this work, we conjecture that they are related to the lattice corrections to the entropy discussed in detail in Ref. \cite{fc-11}.

\subsection{Negativity}

In this section, we review the methods to compute negativity in the lattice, following closely Ref. \cite{cmc-22}. We employ the notion of fermionic negativity, defined in \cite{rac-16b,ssr-17,srrc-19,sr-19,sr-19a,srgr-18}, and related to a notion of \textit{time-reversal transposition}, that is naturally suited for free fermions.

Let us consider a tripartition $A\cup B \cup \overline{(A\cup B)}$. We first define the matrix $\Gamma = 1-2C$, dubbed as \textit{covariance matrix}, we restrict it to $A \cup B$ and we call $\Gamma_{A\cup B}$ the restriction. The latter has a natural block structure
\be
\Gamma_{A\cup B}=  \begin{pmatrix} \Gamma_{AA} & \Gamma_{AB}  \\ \Gamma_{BA}  & \Gamma_{BB}\end{pmatrix},
\ee
where the blocks encode the correlation between the subsystems $A$ and $B$.
We construct the matrix
\be
\Gamma^\times_{A\cup B}\equiv \frac{2}{1+\Gamma^2_{A \cup B}} \begin{pmatrix} -\Gamma_{AA} & 0 \\ 0 & \Gamma_{BB}\end{pmatrix},
\label{eq:GammaxAB}
\ee
and, eventually, we compute the R\'enyi negativity as
\begin{align}
\nonumber
&\mathcal{E}_{n_e} \equiv \log \text{Tr}\l |\rho_{A\cup B}|^{n_e} \r = \text{Tr} \log\l \l\frac{1+\Gamma^\times_{A\cup B}}{2}\r^{n_e/2} + \l\frac{1-\Gamma^\times_{A\cup B}}{2}\r^{n_e/2} \r +\\ 
 &\frac{n_e}{2}\text{Tr} \log\l \l\frac{1+\Gamma_{A\cup B}}{2}\r^{2} + \l\frac{1-\Gamma_{A\cup B}}{2}\r^{2} \r.
\end{align}
\label{eq:NegDefect}
The limit $n_e \rightarrow 1$ of Eq. \eqref{eq:NegDefect} gives the logarithmic negativity between $A$ and $B$, denoted by $\mathcal{E}$.

\begin{figure}[thp]
    \centering
    \includegraphics[width=0.6\textwidth]{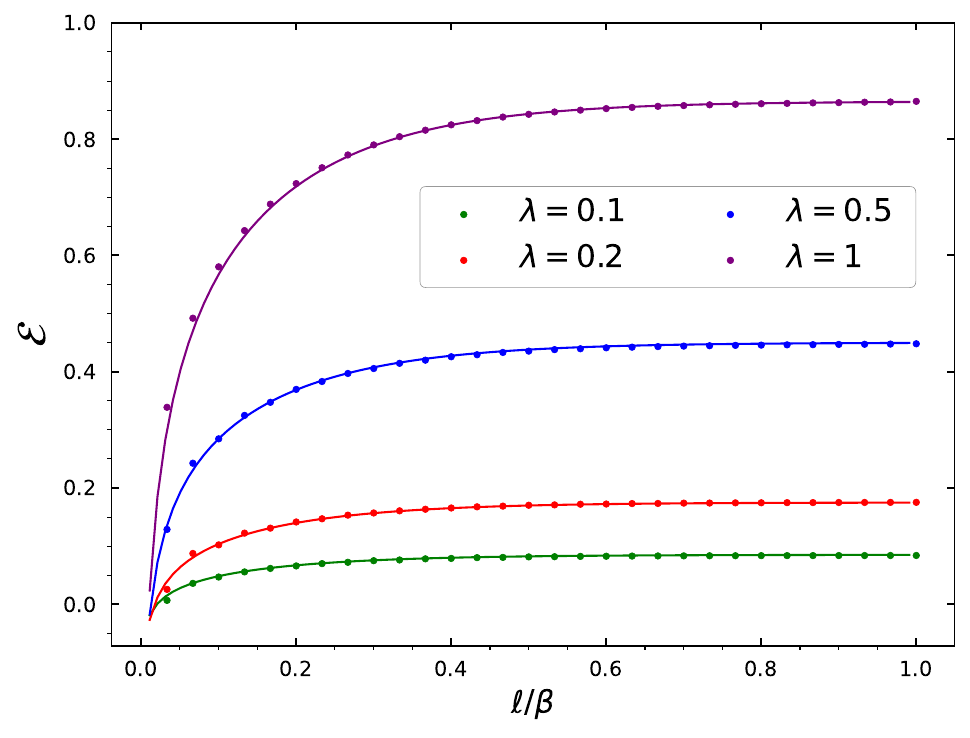}
    \caption{Logarithmic negativity $\mathcal{E}$, as a function of $\ell/\beta$, between two intervals of length $\ell$ attached to the defect. Some values of the transmission amplitude have been considered ($\lambda=0.1,0.2,0.5,1.0$). A fast exponential saturation to a non-universal value (that has been fitted) is observed, while the exact crossover is predicted analytically from Eq. \eqref{eq:Anal_E}.}
    \label{fig:E}
\end{figure}

We consider the system of Sec, \eqref{sec:Num_Renyi}, namely we take two wires ($M=2$) and the scattering matrix $S$ in Eq. \eqref{eq:Scatt_Mat}. We consider $A$ and $B$ as two intervals of length $\ell$ attached to the defect belonging to the first/second wire respectively. Specifically, the sites of $A$ are $x \in [1/2,\dots,\ell-1/2],j=1$ while the sites of $B$ are $x \in [1/2,\dots,\ell-1/2],j=2$. We need to specify the value of $\hat{\Delta}_b$ appearing in the analytical prediction Eq. \eqref{eq:Anal_E}. The latter gives the logarithmic prefactor of the ground state negativity that has been reported in \cite{ge-20}
\be
\underset{n_e\rightarrow 1}{\lim}\hat{\Delta}_b = \frac{1}{\pi^2}\arcsin(\lambda)\l \pi-\arcsin(\lambda)\r.
\ee
We numerically compute the thermal state's negativity with $L=30,\beta = 30$ as a function of $\ell$. Our data are shown in Fig. \eqref{fig:E}, and they are compared with the analytical predictions. The matching is good, even for $\ell \simeq L$ where finite-size effects might play a role, in principle. Our physical understanding is that, for $\ell/\beta$ sufficiently large, the negativity should depend solely on the quantum correlations across the defect, where $A$ and $B$ touch. No contributions from other entangling points should matter. In particular, we observe no UV divergence when the point $x=\ell$ approaches the other boundary at $x=L$.

\section{Conclusions}\label{sec:Conclusions}

We have developed a formalism to characterize the entanglement content of interfaces at finite temperature. Our approach relies on the relation between entanglement measures and a family of bulk/boundary twist fields with non-trivial scaling dimensions. We are able to provide exact results in specific cases, based on conformal invariance only, and we also discuss the relevant physics which appears at small/high temperatures. We show that while at large distances the entropy is always extensive, and it is not affected by the defect (at leading order), both the mutual information and the negativity of intervals attached to the defect satisfy area law. In particular, their value diverges logarithmically as the temperature is decreased, with a universal prefactor related to the boundary scaling dimensions of the twist fields.

Some questions are left open. For example, on the technical side, it would be interesting to characterize analytically multipoint-correlation functions of bulk and boundary twist fields. To the best of our knowledge, even in the case of free theories with many species and mixing boundary conditions (say, a non-trivial boundary scattering matrix) this task has not been accomplished. In principle, one could try to reformulate the boundary CFT problem as a chiral CFT via an unfolding map \cite{dms-97}, where the latter depends on the boundary condition, and then adapt the techniques of Dixon et al. of orbifold CFT \cite{dfms-87}. However, the main issue would be the presence of non-trivial monodromies induced by the twists, and a non-abelian Riemann-Hilbert problem would appear (see Ref. \cite{mf-23}). It is not clear to us if explicit analytical results can be obtained within this approach.
Also, a deeper understanding of the boundary state ${}_{\beta}\bra{b}\sigma_b(0)$, introduced in Eq. \eqref{eq:BState_sigma}, is preferable. In principle, since this is a state of the twisted Hamiltonian, it should be expressed as a linear combination of eigenstates of the latter. Still, it is not obvious whether it is non-normalizable (as it happens for the Ishibashi states \cite{dms-97}), and which eigenstates should appear in its expansion.

Our methods have the potential to be adapted for addressing topological interfaces \cite{gm-16,rs-22,rpr-22,hfsr-23}. In such instances, defects exhibit purely transmissive behavior, leading to intriguing boundary effects governed by the Affleck-Ludwig term. These effects are linked to the universal structure constant, which we previously overlooked, as they manifest as $O(1)$ terms that were not thoroughly discussed. Similarly, one should be capable of modifying the formalism to characterize excited states, previously investigated in \cite{txas-13} for BCFT, in order to replicate the findings of \cite{sths-22,ce-23a}.
It would be of interest to explore whether relevant defects can be examined similarly. In such cases, the boundary conditions lack scale invariance, introducing a non-trivial renormalization group flow. While perturbation theory suffices for short distances, a systematic approach to addressing large distances remains unclear (see Ref. \cite{kss-21}).

Another possible application is the effect of symmetry-breaking boundary conditions in CFT for thermal states, extending the results of Ref. \cite{bcp-23} of the ground state. We expect an exponential decay or saturation of the string-order parameter, as a consequence of a finite correlation length, and the results of Sec. \ref{sec:CFT_half_line} should apply directly.

An interesting question, that remains so far unsolved, is the behavior of the entropy of a finite interval attached to the defect after a global quench. One expects that, for large enough time, the stationary state is reached, and, according to the results of this work, together with the general analysis of \cite{bf-16} for lattice free fermions (see also Refs. \cite{lsp-19}), the information of the defect should be lost. On the other end at short time a linear growth should be observed, and, from the results of \cite{ce-23,wwr-17}, the rate should be proportional to $c_{\text{eff}}+c$. It is not clear to us how these two regimes can be interpolated. A scenario has been recently proposed in Ref. \cite{kkosw-23}, where the authors conjecture that the entropy should be proportional to $c_{\text{eff}}+c$ at large times: to us, this conjecture looks unphysical, and incompatible with the results of this and previous works as well. We believe that, by adapting the results of \cite{ce-23}, and performing a careful analysis of a two-point function of twist fields, one should be able to tackle the problem. 

\medskip {\bf Acknowledgements:} 
LC acknowledges support from ERC under Consolidator grant number 771536 (NEMO), and under Starting grant 805252 LoCoMacro. LC thanks LPTHE (Sorbonne, Paris) for hospitality during a one-month visit in March 2023. The authors are both grateful to Benoit Estienne for stimulating discussion on the topic and insightful comments at the early stages of this project.

\noindent

\begin{appendices}

\section{The boundary scaling dimension of twist fields}

\begin{figure}[t]
    \centering
\includegraphics[width=0.4\linewidth]{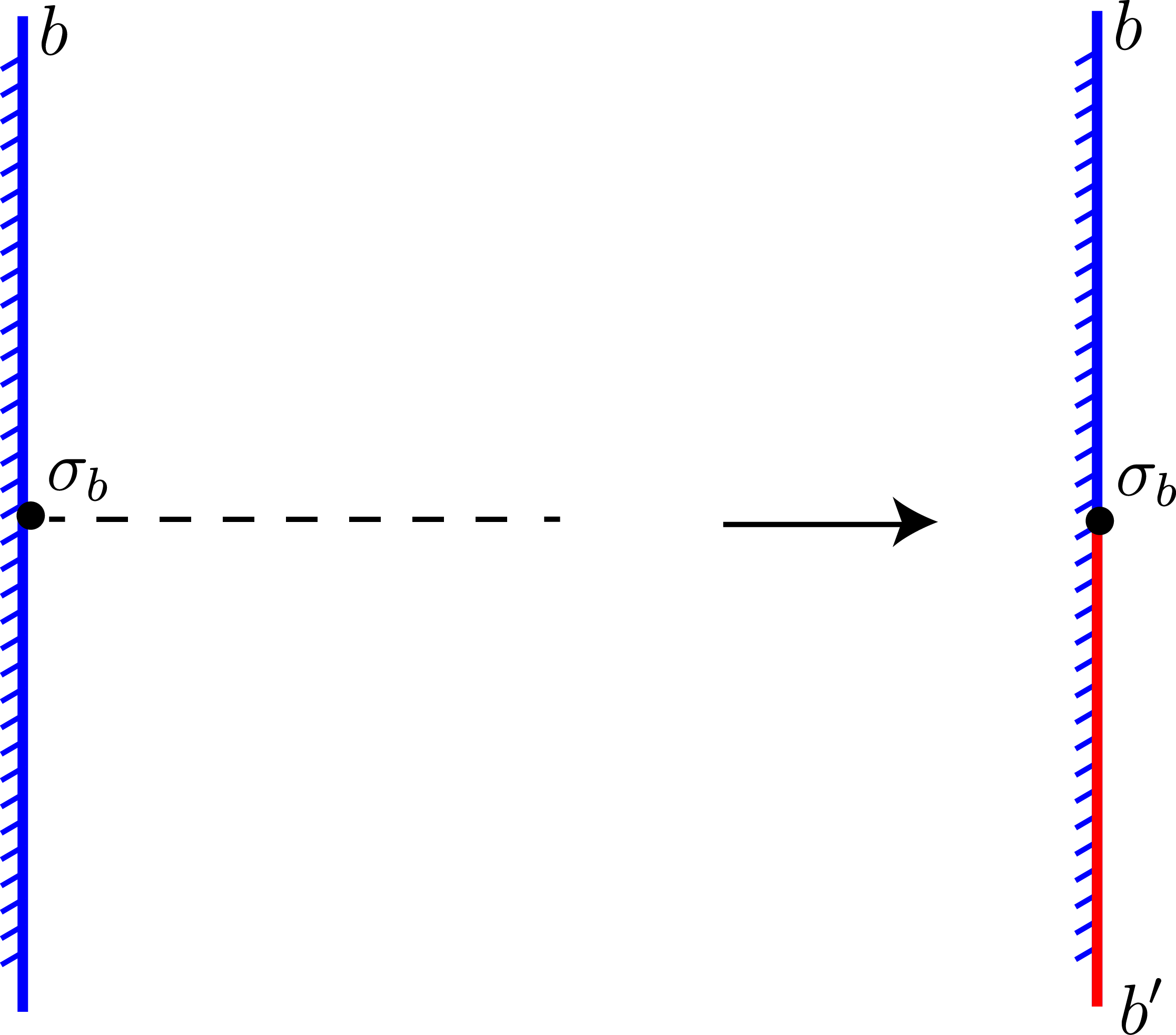}
    \caption{ (Left) Insertion of the boundary twist field $\sigma_b$ in the geometry $\text{Re}(z)\geq 0$, with boundary conditions $b$ at $\text{Re}(z)= 0$. (Right) Boundary changing operator interpolating the boundary condition $b'$, for $z \in -i(0,\infty)$, and $b$, for $z \in +i(0,\infty)$.}
    \label{fig:B_changing}

    \medskip
    \includegraphics[width=0.6\linewidth]{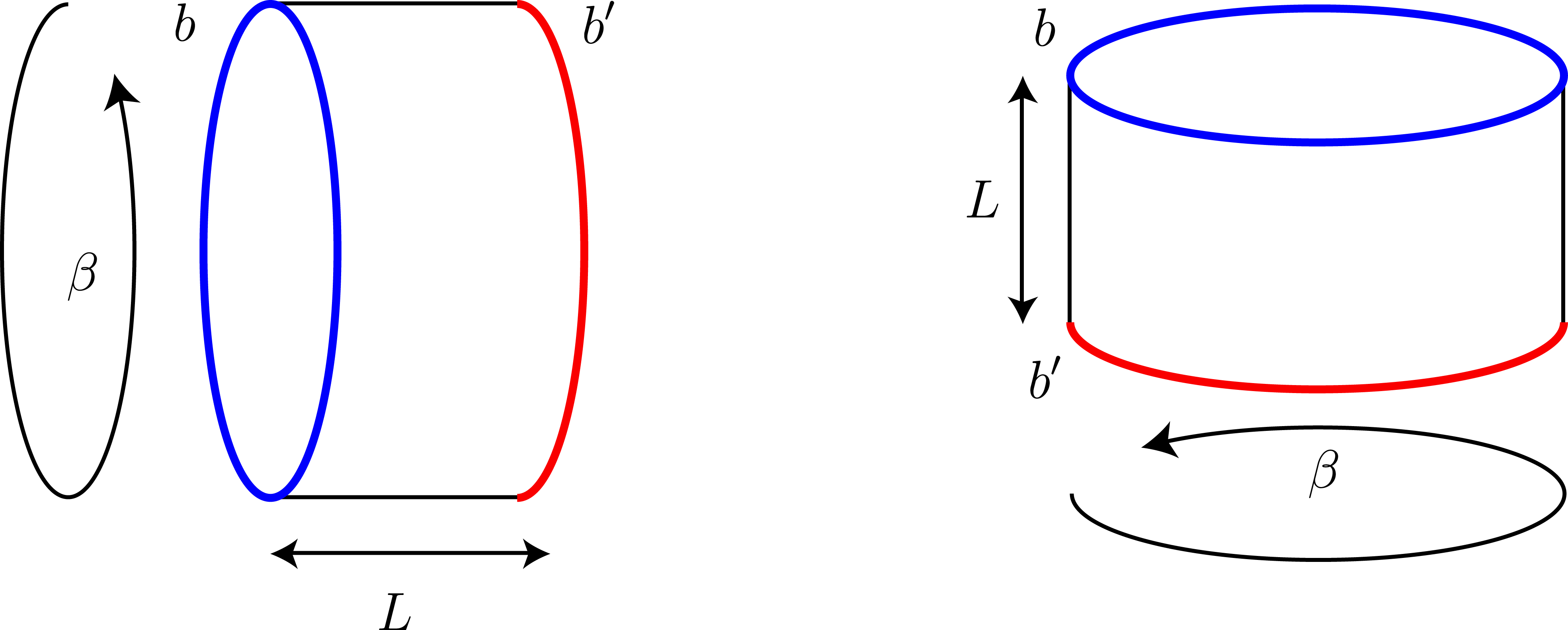}
    \caption{ (Left) Partition function $\text{Tr}\l \exp\l -\beta H_{bb'}\r\r$, with boundary conditions $b,b'$ chosen at the spatial points $z=0,L$. (Right) Partition function $\bra{b}\exp \l-LH \r\ket{b'}$ in the crossed channel, expressed as an overlap between boundary states.}
    \label{fig:Crossed}
    
\end{figure}

In this appendix, we give a detailed description of the boundary scaling dimensions of the twist fields, following standard techniques of BCFT \cite{dms-97,Recknagel:2013uja}. In particular, we first relate them to the ground state energy of a specific finite-size Hamiltonian. Then, via modularity, we show that it corresponds to a thermal free energy density, where non-trivial boundary conditions are kept along the (Euclidean) time direction.
Let us consider the geometry
\be
\text{Re}(z)\geq 0,
\ee
where boundary conditions of type $b$ are inserted at $\text{Re}(z)= 0$. We insert the twist field $\sigma$ at the boundary point $z=0$, and we denote it by $\sigma_b(0)$. A branch-cut originates from the boundary point $z=0$ and propagates along the half-line $z \in [0,\infty)$: we deform it, until it propagates on the imaginary axis $z \in -i[0,\infty)$. According to this procedure, a new boundary condition is generated on
\be
z \in -i[0,\infty),
\ee
and we denote it by $b'$. In this way, a sudden change of boundary condition (from $b'$ to $b$), occurs at $z=0$, and therefore $\sigma_b(0)$ plays the role of a boundary condition changing operator. We summarize the procedure above in Fig. \ref{fig:B_changing}.

We map to the strip
\be
\text{Re}(\zeta) \in [-L/2,L/2]
\ee
via the transformation $z = \exp\l -\frac{i\pi \zeta}{L}\r$. The boundary conditions at $\text{Re}(\zeta) = \mp L/2$ are $b,b'$ respectively. The resulting geometry gives the zero-temperature phase of a finite-size Hamiltonian $H_{bb'}$ defined on a segment of length $L$. One can show via BCFT techniques that $H_{bb'}$ is described by a chiral CFT \cite{dms-97,Cardy-04}, and it holds
\be\label{eq:chiral_Ham}
H_{bb'} = \frac{\pi}{L}\l L_0 - \frac{c}{24} \r,
\ee
where the dependence on the boundary conditions $b,b'$ enters the specific representation of the Virasoro algebra. In particular, the ground state energies of $H_{bb}$ and $H_{bb'}$ are different in principle.
Via state-operator correspondence, the lightest state of $H_{bb'}$, which we denote $|0\rangle_{bb'}$, corresponds precisely to the boundary-changing operator. Moreover, the difference of the ground state energies of $H_{bb'}$ and $H_{bb}$ is precisely 
\be\label{eq:GS_en_diff}
E_{\text{g.s.},bb'} - E_{\text{g.s.},bb} = \frac{\pi \Delta_b}{L}.
\ee
From \eqref{eq:GS_en_diff} we learn that a non-trivial boundary scaling dimension $\Delta_b$ of the twist field $\sigma$ can be ultimately related to the energy cost to change of boundary conditions from $b$ to $b'$.

Thanks to \eqref{eq:GS_en_diff}, we now express the low-temperature limit of the ratio of partition functions as
\be\label{eq:Hbb1}
\frac{\text{Tr}\l \exp\l -\beta H_{bb'}\r\r}{\text{Tr}\l \exp\l -\beta H_{bb}\r\r} \sim \exp\l -\frac{\pi \beta}{L} \Delta_b\r, \quad \beta \gg L,
\ee
By exploiting the world-sheet duality of the partition function, we  can equivalently obtain it by quantizing the system in the crossed channel where time and space are exchanged. We do so, and we end up with the Hamiltonian
\be
H = \frac{2\pi}{\beta}\l L_0 + \bar{L}_0 - \frac{c}{12} \r,
\ee
with periodic boundary conditions along the circle of length $\beta$. In this picture, the boundary conditions $b$ and $b'$ are fixed along the space direction, and they are expressed by some boundary states $\ket{b},\ket{b'}$. Finally, we express the partition function as
\be\label{eq:B_overlap}
\text{Tr}\l \exp\l -\beta H_{bb'}\r\r = \bra{b}\exp \l-LH \r\ket{b'},
\ee
and similarly
\be
\text{Tr}\l \exp\l -\beta H_{bb}\r\r = \bra{b}\exp \l-LH \r\ket{b}.
\ee

In the limit $\beta\gg L$, corresponding to the large subsystem size in the crossed channel, the free energy becomes extensive (in $\beta$), and we have
\be\label{eq:crossed}
\bra{b}\exp \l-LH \r\ket{b'} \sim \exp\l -L\beta \times \text{"density of free energy"}\r.
\ee
Comparing the result \eqref{eq:Hbb1} with Eq. \eqref{eq:crossed}, we obtain the exact value of the difference of free energy density as
\be\label{eq:free_en}
\underset{\beta \rightarrow +\infty}{\lim}-\frac{1}{\beta L} \log \frac{\bra{b}\exp \l-LH \r\ket{b'}}{\bra{b}\exp \l-LH \r\ket{b}} = \frac{\pi \Delta_b}{L^2}.
\ee
Together with \eqref{eq:GS_en_diff}, Eq. \eqref{eq:free_en} is the main result of this appendix, and it gives a physical interpretation of the boundary scaling dimension $\Delta_b$ of the twist field $\sigma$.
We mention that detailed calculations of the overlap \eqref{eq:B_overlap} have been performed in \cite{ss-08,bb-15,gm-17,cmc-22,cmc-22a} for free theories to characterize the ground state entanglement of conformal junctions. However, their relation with the boundary twist fields is, up to our knowledge, new, albeit it follows directly from standard BCFT arguments.

A final technical remark is needed. To derive Eq. \eqref{eq:GS_en_diff}, we assumed implicitly that $\sigma_b$ is the lightest operator interpolating the boundary conditions $b$ and $b'$. Still, an infinite tower of excited states above the ground state is present for $H_{bb'}$ \eqref{eq:chiral_Ham}. For example, if one fuses the twist field $\sigma$ with a local operator $\mathcal{O}$, the resulting composite field $(\sigma\cdot \mathcal{O})$ would have the same semilocality properties of $\sigma$; therefore, the boundary operator obtained from the fusion of $(\sigma\cdot \mathcal{O})$ and the boundary $b$, has to be interpreted as a boundary changing operator between $b$ and $b'$. In particular, by state-operator correspondence, it corresponds to a state of the Hamiltonian $H_{bb'}$.

\end{appendices}

\printbibliography

@article{Dupic:2017hpb,
    author = "Dupic, Thomas and Estienne, Benoit and Ikhlef, Yacine",
    title = "{Entanglement entropies of minimal models from null-vectors}",
    eprint = "1709.09270",
    archivePrefix = "arXiv",
    primaryClass = "math-ph",
    doi = "10.21468/SciPostPhys.4.6.031",
    journal = "SciPost Phys.",
    volume = "4",
    number = "6",
    pages = "031",
    year = "2018"
}

@article{Mintchev:2020jhc,
    author = "Mintchev, Mihail and Tonni, Erik",
    title = "{Modular Hamiltonians for the massless Dirac field in the presence of a defect}",
    eprint = "2012.01366",
    archivePrefix = "arXiv",
    primaryClass = "hep-th",
    doi = "10.1007/JHEP03(2021)205",
    journal = "JHEP",
    volume = "03",
    pages = "205",
    year = "2021"
}

@phdthesis{runkel2000boundary,
  title={Boundary problems in conformal field theory},
  author={Runkel, Ingo},
  year={2000},
  school={University of London}
}

@book{Recknagel:2013uja,
    author = {Recknagel, Andreas and Schomerus, Volker},
    title = "{Boundary Conformal Field Theory and the Worldsheet Approach to D-Branes}",
    doi = {10.1017/CBO9780511806476},
    isbn = "978-0-521-83223-6, 978-0-521-83223-6, 978-1-107-49612-5",
    publisher = "Cambridge University Press",
    series = "Cambridge Monographs on Mathematical Physics",
    month = "11",
    year = "2013"
}

@book{dms-97,
    author = "Di Francesco, P. and Mathieu, P. and Senechal, D.",
    title = "{Conformal Field Theory}",
    doi = "10.1007/978-1-4612-2256-9",
    publisher = "Springer-Verlag",
    address = "New York",
    series = "Graduate Texts in Contemporary Physics",
    year = "1997"
}

@article{Cardy-84,
    author = "Cardy, John L.",
    title = "{Conformal Invariance and Surface Critical Behavior}",
    doi = "10.1016/0550-3213(84)90241-4",
    journal = "Nucl. Phys. B",
    volume = "240",
    pages = "514--532",
    year = "1984"
}

@article{Cardy-04,
  title={Boundary Conformal Field Theory},
  author={John Cardy},
  journal={arXiv: High Energy Physics - Theory},
  year={2004},
  url={https://api.semanticscholar.org/CorpusID:15725963}
}

@article{al-91a,
  author = "Affleck, Ian and Ludwig, Andreas W. W.",
    title = "{Universal noninteger 'ground state degeneracy' in critical quantum systems}",
    reportNumber = "UBCTP-91-007",
    doi = "10.1103/PhysRevLett.67.161",
    journal = "Phys. Rev. Lett.",
    volume = "67",
    pages = "161--164",
    year = "1991"
}

@article{wvhc-08,
  author = "Wolf, Michael M. and Verstraete, Frank and Hastings, Matthew B. and Cirac, J. Ignacio",
    title = "{Area Laws in Quantum Systems: Mutual Information and Correlations}",
    eprint = "0704.3906",
    archivePrefix = "arXiv",
    primaryClass = "quant-ph",
    doi = "10.1103/PhysRevLett.100.070502",
    journal = "Phys. Rev. Lett.",
    volume = "100",
    number = "7",
    pages = "070502",
    year = "2008"
}

@ARTICLE{Kondo-64,
       author = {{Kondo}, J.},
        title = "{Resistance Minimum in Dilute Magnetic Alloys}",
      journal = {Progress of Theoretical Physics},
         year = 1964,
        month = jul,
       volume = {32},
       number = {1},
        pages = {37-49},
          doi = {10.1143/PTP.32.37},
       adsurl = {https://ui.adsabs.harvard.edu/abs/1964PThPh..32...37K}
}

@article{kf-92,
  title = {Transmission through barriers and resonant tunneling in an interacting one-dimensional electron gas},
  author = {Kane, C. L. and Fisher, Matthew P. A.},
  journal = {Phys. Rev. B},
  volume = {46},
  issue = {23},
  pages = {15233--15262},
  numpages = {0},
  year = {1992},
  publisher = {American Physical Society},
  doi = {10.1103/PhysRevB.46.15233},
  url = {https://link.aps.org/doi/10.1103/PhysRevB.46.15233}
}

@article{al-91,
  author = "Affleck, Ian and Ludwig, Andreas W. W.",
    title = "{Critical theory of overscreened Kondo fixed points}",
    reportNumber = "UBCTP-90-7",
    doi = "10.1016/0550-3213(91)90419-X",
    journal = "Nucl. Phys. B",
    volume = "360",
    pages = "641--696",
    year = "1991"
}

@article{oa-96,
    author = "Oshikawa, Masaki and Affleck, Ian",
    title = "{Defect lines in the Ising model and boundary states on orbifolds}",
    eprint = "hep-th/9606177",
    archivePrefix = "arXiv",
    doi = "10.1103/PhysRevLett.77.2604",
    journal = "Phys. Rev. Lett.",
    volume = "77",
    pages = "2604--2607",
    year = "1996"
}

@article{al-93,
  title = {{Exact conformal-field-theory results on the multichannel Kondo effect: Single-fermion Green's function, self-energy, and resistivity}},
  author = {Affleck, Ian and Ludwig, Andreas W. W.},
  journal = {Phys. Rev. B},
  volume = {48},
  issue = {10},
  pages = {7297--7321},
  numpages = {0},
  year = {1993},
  publisher = {American Physical Society},
  doi = {10.1103/PhysRevB.48.7297},
  url = {https://link.aps.org/doi/10.1103/PhysRevB.48.7297}
}

@article{kss-21,
  title = {{Universal Thermal Entanglement of Multichannel Kondo Effects}},
  author = {Kim, Donghoon and Shim, Jeongmin and Sim, H.-S.},
  journal = {Phys. Rev. Lett.},
  volume = {127},
  issue = {22},
  pages = {226801},
  numpages = {7},
  year = {2021},
  publisher = {American Physical Society},
  doi = {10.1103/PhysRevLett.127.226801},
  url = {https://link.aps.org/doi/10.1103/PhysRevLett.127.226801}
}

@article{Plenio-05,
  title={Logarithmic negativity: a full entanglement monotone that is not convex},
  author={Plenio, Martin B},
  journal={PRL},
  volume={95},
  number={9},
  pages={090503},
  year={2005},
  publisher={APS},
  doi = {10.1103/PhysRevLett.95.090503}
}

@article{ss-08,
    author = "Sakai, Kazuhiro and Satoh, Yuji",
    title = "{Entanglement through conformal interfaces}",
    eprint = "0809.4548",
    archivePrefix = "arXiv",
    primaryClass = "hep-th",
    reportNumber = "UTHEP-570",
    doi = "10.1088/1126-6708/2008/12/001",
    journal = "JHEP",
    volume = "12",
    pages = "001",
    year = "2008"
}

@article{Peschel-05,
title={Entanglement entropy with interface defects},
   volume={38},
   ISSN={1361-6447},
   url={http://dx.doi.org/10.1088/0305-4470/38/20/002},
   DOI={10.1088/0305-4470/38/20/002},
   number={20},
   journal={J. Phys. A: Math. Gen.},
   publisher={IOP Publishing},
   author={Peschel, Ingo},
   year={2005},
   month=may, pages={4327–4335}
}

@article{cc-04,
    author = "Calabrese, Pasquale and Cardy, John L.",
    title = "{Entanglement entropy and quantum field theory}",
    eprint = "hep-th/0405152",
    archivePrefix = "arXiv",
    doi = "10.1088/1742-5468/2004/06/P06002",
    journal = "J. Stat. Mech.",
    volume = "0406",
    pages = "P06002",
    year = "2004"
}

@article{cc-09,
    author = "Calabrese, Pasquale and Cardy, John",
    title = "{Entanglement entropy and conformal field theory}",
    eprint = "0905.4013",
    archivePrefix = "arXiv",
    primaryClass = "cond-mat.stat-mech",
    doi = "10.1088/1751-8113/42/50/504005",
    journal = "J. Phys. A",
    volume = "42",
    pages = "504005",
    year = "2009"
}

@Article{ep-10,
  title =        {Entanglement in fermionic chains with interface defects},
  author =       {V. Eisler and I. Peschel},
  journal =      {Ann. Phys. (Berlin)},
  year =         {2010},
  volume =       {522},
  pages =        {679},
doi="10.1002/andp.201000055"
}

@article{isl-09,
  title={Entanglement entropy with localized and extended interface defects},
  author={Igl{\'o}i, Ferenc and Szatm{\'a}ri, Zsolt and Lin, Yu-Cheng},
  journal={Physical Review B},
  volume={80},
  number={2},
  pages={024405},
  year={2009},
  publisher={APS},
doi="10.1103/PhysRevB.80.024405
Focus to learn more
"
}

@article{cmv-12,
    author = "Calabrese, Pasquale and Mintchev, Mihail and Vicari, Ettore",
    title = "{Entanglement Entropy of Quantum Wire Junctions}",
    eprint = "1110.5713",
    archivePrefix = "arXiv",
    primaryClass = "cond-mat.stat-mech",
    reportNumber = "IFUP-TH-21-2011",
    doi = "10.1088/1751-8113/45/10/105206",
    journal = "J. Phys. A",
    volume = "45",
    pages = "105206",
    year = "2012"
}

@article{bb-15,
    author = "Brehm, Enrico M. and Brunner, Ilka",
    title = "{Entanglement entropy through conformal interfaces in the 2D Ising model}",
    eprint = "1505.02647",
    archivePrefix = "arXiv",
    primaryClass = "hep-th",
    reportNumber = "LMU-ASC-23-15",
    doi = "10.1007/JHEP09(2015)080",
    journal = "JHEP",
    volume = "09",
    pages = "080",
    year = "2015"
}

@article{gm-17,
     author = "Gutperle, Michael and Miller, John D.",
    title = "{Entanglement entropy at CFT junctions}",
    eprint = "1701.08856",
    archivePrefix = "arXiv",
    primaryClass = "hep-th",
    doi = "10.1103/PhysRevD.95.106008",
    journal = "Phys. Rev. D",
    volume = "95",
    number = "10",
    pages = "106008",
    year = "2017"
}

@article{cmc-22,
    author = "Capizzi, Luca and Murciano, Sara and Calabrese, Pasquale",
    title = "{R\'enyi entropy and negativity for massless Dirac fermions at conformal interfaces and junctions}",
    eprint = "2205.04722",
    archivePrefix = "arXiv",
    primaryClass = "hep-th",
    doi = "10.1007/JHEP08(2022)171",
    journal = "JHEP",
    volume = "08",
    pages = "171",
    year = "2022"
}

@article{cmc-22a,
    author = "Capizzi, Luca and Murciano, Sara and Calabrese, Pasquale",
    title = "{R\'enyi entropy and negativity for massless complex boson at conformal interfaces and junctions}",
    eprint = "2208.14118",
    archivePrefix = "arXiv",
    primaryClass = "hep-th",
    doi = "10.1007/JHEP11(2022)105",
    journal = "JHEP",
    volume = "11",
    pages = "105",
    year = "2022"
}

@article{ce-23,
    author = "Capizzi, Luca and Eisler, Viktor",
    title = "{Entanglement evolution after a global quench across a conformal defect}",
    eprint = "2209.03297",
    archivePrefix = "arXiv",
    primaryClass = "cond-mat.stat-mech",
    doi = "10.21468/SciPostPhys.14.4.070",
    journal = "SciPost Phys.",
    volume = "14",
    number = "4",
    pages = "070",
    year = "2023"
}

@article{wwr-17,
author = {Wen, Xueda and Wang, Yuxuan and Ryu, Shinsei},
year = {2018},
month = {11},
pages = {},
title = {Entanglement evolution across a conformal interface},
volume = {51},
journal = {J. Phys. A Math. Theor.},
doi = {10.1088/1751-8121/aab561}
}

@article{bd-17,
 author = "Blondeau-Fournier, Olivier and Doyon, Benjamin",
    title = "{Expectation values of twist fields and universal entanglement saturation of the free massive boson}",
    eprint = "1612.04238",
    archivePrefix = "arXiv",
    primaryClass = "hep-th",
    doi = "10.1088/1751-8121/aa7492",
    journal = "J. Phys. A",
    volume = "50",
    number = "27",
    pages = "274001",
    year = "2017"
}

@article{hn-13,
   author = "Herzog, Christopher P. and Nishioka, Tatsuma",
    title = "{Entanglement Entropy of a Massive Fermion on a Torus}",
    eprint = "1301.0336",
    archivePrefix = "arXiv",
    primaryClass = "hep-th",
    reportNumber = "PUPT-2438, YITP-12-47",
    doi = "10.1007/JHEP03(2013)077",
    journal = "JHEP",
    volume = "03",
    pages = "077",
    year = "2013"
}

@article{dd-14,
    author = "Datta, Shouvik and David, Justin R.",
    title = "{R\'enyi entropies of free bosons on the torus and holography}",
    eprint = "1311.1218",
    archivePrefix = "arXiv",
    primaryClass = "hep-th",
    doi = "10.1007/JHEP04(2014)081",
    journal = "JHEP",
    volume = "04",
    pages = "081",
    year = "2014"
}

@article{cw-14,
  author = "Chen, Bin and Wu, Jie-qiang",
    title = "{Single interval Renyi entropy at low temperature}",
    eprint = "1405.6254",
    archivePrefix = "arXiv",
    primaryClass = "hep-th",
    doi = "10.1007/JHEP08(2014)032",
    journal = "JHEP",
    volume = "08",
    pages = "032",
    year = "2014"
}

@article{ch-14,
    author = "Cardy, John and Herzog, Christopher P.",
    title = "{Universal Thermal Corrections to Single Interval Entanglement Entropy for Two Dimensional Conformal Field Theories}",
    eprint = "1403.0578",
    archivePrefix = "arXiv",
    primaryClass = "hep-th",
    doi = "10.1103/PhysRevLett.112.171603",
    journal = "Phys. Rev. Lett.",
    volume = "112",
    number = "17",
    pages = "171603",
    year = "2014"
}

@article{hn-15,
    author = "Herzog, Christopher P. and Nian, Jun",
    title = "{Thermal corrections to R\'enyi entropies for conformal field theories}",
    eprint = "1411.6505",
    archivePrefix = "arXiv",
    primaryClass = "hep-th",
    reportNumber = "YITP-SB-14-48, YITP-SB-14-48",
    doi = "10.1007/JHEP06(2015)009",
    journal = "JHEP",
    volume = "06",
    pages = "009",
    year = "2015"
}

@article{cct-14,
   author = "Calabrese, Pasquale and Cardy, John and Tonni, Erik",
    title = "{Finite temperature entanglement negativity in conformal field theory}",
    eprint = "1408.3043",
    archivePrefix = "arXiv",
    primaryClass = "cond-mat.stat-mech",
    doi = "10.1088/1751-8113/48/1/015006",
    journal = "J. Phys. A",
    volume = "48",
    number = "1",
    pages = "015006",
    year = "2015"
}

@article{cw-15,
  author = "Chen, Bin and Wu, Jie-qiang",
    title = "{Large interval limit of R\'enyi entropy at high temperature}",
    eprint = "1412.0763",
    archivePrefix = "arXiv",
    primaryClass = "hep-th",
    doi = "10.1103/PhysRevD.92.126002",
    journal = "Phys. Rev. D",
    volume = "92",
    number = "12",
    pages = "126002",
    year = "2015"
}

@article{hs-16,
     author = "Herzog, Christopher P. and Spillane, Michael",
    title = "{Thermal corrections to R\'enyi entropies for free fermions}",
    eprint = "1506.06757",
    archivePrefix = "arXiv",
    primaryClass = "hep-th",
    doi = "10.1007/JHEP04(2016)124",
    journal = "JHEP",
    volume = "04",
    pages = "124",
    year = "2016"
}

@article{sr-19,
      author = "Shapourian, Hassan and Ryu, Shinsei",
    title = "{Finite-temperature entanglement negativity of free fermions}",
    eprint = "1807.09808",
    archivePrefix = "arXiv",
    primaryClass = "cond-mat.stat-mech",
    doi = "10.1088/1742-5468/ab11e0",
    journal = "J. Stat. Mech.",
    volume = "1904",
    pages = "043106",
    year = "2019"
}

@article{wlckg-20,
author = "Wu, Kai-Hsin and Lu, Tsung-Cheng and Chung, Chia-Min and Kao, Ying-Jer and Grover, Tarun",
    title = "{Entanglement Renyi Negativity across a Finite Temperature Transition: A Monte Carlo study}",
    eprint = "1912.03313",
    archivePrefix = "arXiv",
    primaryClass = "cond-mat.str-el",
    doi = "10.1103/PhysRevLett.125.140603",
    journal = "Phys. Rev. Lett.",
    volume = "125",
    number = "14",
    pages = "140603",
    year = "2020"
}

@article{rmc-23,
    author = "Rottoli, Federico and Murciano, Sara and Calabrese, Pasquale",
    title = "{Finite temperature negativity Hamiltonians of the massless Dirac fermion}",
    eprint = "2304.09906",
    archivePrefix = "arXiv",
    primaryClass = "hep-th",
    doi = "10.1007/JHEP06(2023)139",
    journal = "JHEP",
    volume = "06",
    pages = "139",
    year = "2023"
}

@article{bs-06,
  title="Boundary effects in the critical scaling of entanglement entropy in 1D systems",
  author={Laflorencie, Nicolas and S{\o}rensen, Erik S and Chang, Ming-Shyang and Affleck, Ian},
  journal={Phys. Rev. Lett.},
  volume={96},
  number={10},
  pages={100603},
  year={2006},
  publisher={APS},
 doi="10.1103/PhysRevLett.96.100603"
}

@article{bs-16,
     author = "Berthiere, Clement and Solodukhin, Sergey N.",
    title = "{Boundary effects in entanglement entropy}",
    eprint = "1604.07571",
    archivePrefix = "arXiv",
    primaryClass = "hep-th",
    doi = "10.1016/j.nuclphysb.2016.07.029",
    journal = "Nucl. Phys. B",
    volume = "910",
    pages = "823--841",
    year = "2016"
}

@article{als-09,
  title={Entanglement entropy in quantum impurity systems and systems with boundaries},
  author={Affleck, Ian and Laflorencie, Nicolas and S{\o}rensen, Erik S},
  journal={Journal of Physics A: Mathematical and Theoretical},
  volume={42},
  number={50},
  pages={504009},
  year={2009},
  publisher={IOP Publishing},
  doi="10.1088/1751-8113/42/50/504009"
}

@article{ntu-12,
 author = "Nozaki, Masahiro and Takayanagi, Tadashi and Ugajin, Tomonori",
    title = "{Central Charges for BCFTs and Holography}",
    eprint = "1205.1573",
    archivePrefix = "arXiv",
    primaryClass = "hep-th",
    reportNumber = "YITP-12-42, IPMU12-0087",
    doi = "10.1007/JHEP06(2012)066",
    journal = "JHEP",
    volume = "06",
    pages = "066",
    year = "2012"
}

@article{Takayanagi-11,
     author = "Takayanagi, Tadashi",
    title = "{Holographic Dual of BCFT}",
    eprint = "1105.5165",
    archivePrefix = "arXiv",
    primaryClass = "hep-th",
    reportNumber = "IPMU11-0091",
    doi = "10.1103/PhysRevLett.107.101602",
    journal = "Phys. Rev. Lett.",
    volume = "107",
    pages = "101602",
    year = "2011"
}

@article{eir-22,
     author = "Estienne, Benoit and Ikhlef, Yacine and Rotaru, Andrei",
    title = "{Second R\'enyi entropy and annulus partition function for one-dimensional quantum critical systems with boundaries}",
    eprint = "2112.01929",
    archivePrefix = "arXiv",
    primaryClass = "math-ph",
    doi = "10.21468/SciPostPhys.12.4.141",
    journal = "SciPost Phys.",
    volume = "12",
    number = "4",
    pages = "141",
    year = "2022"
}

@article{eir-22a,
    author = "Estienne, Benoit and Ikhlef, Yacine and Rotaru, Andrei",
    title = "{The operator algebra of cyclic orbifolds}",
    eprint = "2212.07678",
    archivePrefix = "arXiv",
    primaryClass = "hep-th",
    doi = "10.1088/1751-8121/acfcf6",
    journal = "J. Phys. A",
    volume = "56",
    pages = "465403",
    year = "2023"
}

@article{eir-23,
 author = "Estienne, Benoit and Ikhlef, Yacine and Rotaru, Andrei",
    title = "{R\'enyi entropies for one-dimensional quantum systems with mixed boundary conditions}",
    eprint = "2301.02124",
    archivePrefix = "arXiv",
    journal= " ",
    primaryClass = "cond-mat.stat-mech",
    month = "1",
    year = "2023",
doi=
"10.48550/arXiv.2301.02124
"
}

@article{eirt-23,
  title={Entanglement entropies of an interval for the massless scalar field in the presence of a boundary},
  author={Estienne, Benoit and Ikhlef, Yacine and Rotaru, Andrei and Tonni, Erik},
  journal={ },
  year={2023},
  doi="10.48550/arXiv.2308.00614"
}

@article{txas-13,
  title={Entanglement entropies in conformal systems with boundaries},
  author={Taddia, L and Xavier, JC and Alcaraz, Francisco Castilho and Sierra, G},
  journal={Phys. Rev. B},
  volume={88},
  number={7},
  pages={075112},
  year={2013},
  publisher={APS},
doi="10.1103/PhysRevB.88.075112"
}

@article{mf-23,
  author = "Mari\'c, Vanja and Fagotti, Maurizio",
    title = "{Universality in the tripartite information after global quenches: (generalised) quantum XY models}",
    eprint = "2302.01322",
    archivePrefix = "arXiv",
    primaryClass = "cond-mat.stat-mech",
    doi = "10.1007/JHEP06(2023)140",
    journal = "JHEP",
    volume = "23",
    pages = "140",
    year = "2020"
}

@article{kkosw-23,
  author = "Karch, Andreas and Kusuki, Yuya and Ooguri, Hirosi and Sun, Hao-Yu and Wang, Mianqi",
    title = "{Universality of effective central charge in interface CFTs}",
    eprint = "2308.05436",
    archivePrefix = "arXiv",
    primaryClass = "hep-th",
    reportNumber = "CALT-TH 2023-031, IPMU 23-0025, RIKEN-iTHEMS-Report-23",
    doi = "10.1007/JHEP11(2023)126",
    journal = "JHEP",
    volume = "11",
    pages = "126",
    year = "2023"
}

@article{bf-16,
  title={Determination of the nonequilibrium steady state emerging from a defect},
  author={Bertini, Bruno and Fagotti, Maurizio},
  journal={Phys. Rev. Lett.},
  volume={117},
  number={13},
  pages={130402},
  year={2016},
  publisher={APS},
 doi="https://doi.org/10.1103/PhysRevLett.117.130402"
}

@article{lsp-19,
  title={Non-equilibrium quantum transport in presence of a defect: the non-interacting case},
  author={Ljubotina, Marko and Sotiriadis, Spyros and Prosen, Toma{\v{z}}},
  journal={SciPost Physics},
  volume={6},
  number={1},
  pages={004},
  year={2019},
  doi="https://doi.org/10.21468/SciPostPhys.6.1.004"
}

@article{bcp-23,
     author = "Bonsignori, Riccarda and Capizzi, Luca and Panopoulos, Pantelis",
    title = "{Boundary Symmetry Breaking in CFT and the string order parameter}",
    eprint = "2301.08676",
    archivePrefix = "arXiv",
    primaryClass = "hep-th",
    doi = "10.1007/JHEP05(2023)027",
    journal = "JHEP",
    volume = "05",
    pages = "027",
    year = "2023"
}

@article{gls-23,
  title={Stationary time correlations for fermions after a quench in the presence of an impurity},
  author={Gouraud, Gabriel and Le Doussal, Pierre and Schehr, Gregory},
  journal={Europhysics Letters},
  volume={142},
  number={4},
  pages={41001},
  year={2023},
  publisher={IOP Publishing},
  doi="https://doi.org/10.1209/0295-5075/accec7"
}

@article{csrc-23,
 author = "Capizzi, Luca and Scopa, Stefano and Rottoli, Federico and Calabrese, Pasquale",
    title = "{Domain wall melting across a defect}",
    eprint = "2210.02162",
    archivePrefix = "arXiv",
    primaryClass = "cond-mat.stat-mech",
    doi = "10.1209/0295-5075/acb50a",
    journal = "EPL",
    volume = "141",
    number = "3",
    pages = "31002",
    year = "2023"
}

@article{fg-23,
 author = "Fraenkel, Shachar and Goldstein, Moshe",
    title = "{Extensive long-range entanglement in a nonequilibrium steady state}",
    eprint = "2205.12991",
    archivePrefix = "arXiv",
    primaryClass = "quant-ph",
    doi = "10.21468/SciPostPhys.15.4.134",
    journal = "SciPost Phys.",
    volume = "15",
    number = "4",
    pages = "134",
    year = "2023"
}

@article{gm-16,
author = "Gutperle, Michael and Miller, John D.",
    title = "{A note on entanglement entropy for topological interfaces in RCFTs}",
    eprint = "1512.07241",
    archivePrefix = "arXiv",
    primaryClass = "hep-th",
    doi = "10.1007/JHEP04(2016)176",
    journal = "JHEP",
    volume = "04",
    pages = "176",
    year = "2016"
}

@article{rs-22,
  author = "Roy, Ananda and Saleur, Hubert",
    title = "{Entanglement Entropy in the Ising Model with Topological Defects}",
    eprint = "2111.04534",
    archivePrefix = "arXiv",
    primaryClass = "hep-th",
    doi = "10.1103/PhysRevLett.128.090603",
    journal = "Phys. Rev. Lett.",
    volume = "128",
    number = "9",
    pages = "090603",
    year = "2022"
}

@article{rpr-22,
 author = "Rogerson, David and Pollmann, Frank and Roy, Ananda",
    title = "{Entanglement entropy and negativity in the Ising model with defects}",
    eprint = "2204.03601",
    archivePrefix = "arXiv",
    primaryClass = "hep-th",
    doi = "10.1007/JHEP06(2022)165",
    journal = "JHEP",
    volume = "06",
    pages = "165",
    year = "2022"
}

@article{hfsr-23,
   author = "Horvath, David X. and Fraenkel, Shachar and Scopa, Stefano and Rylands, Colin",
    title = "{Charge-resolved entanglement in the presence of topological defects}",
    eprint = "2306.15532",
    archivePrefix = "arXiv",
    primaryClass = "quant-ph",
    reportNumber = "Phys. Rev. B 108, 165406",
    doi = "10.1103/PhysRevB.108.165406",
    journal = "Phys. Rev. B",
    volume = "108",
    number = "16",
    pages = "165406",
    year = "2023"
}

@article{sths-22,
  	title={{Parity effects and universal terms of $\mathcal{O}(1)$ in the entanglement near a boundary}},
	author={Henning Schlömer and Chunyu Tan and Stephan Haas and Hubert Saleur},
	journal={SciPost Phys.},
	volume={13},
	pages={110},
	year={2022},
	publisher={SciPost},
	doi={10.21468/SciPostPhys.13.5.110},
	url={https://scipost.org/10.21468/SciPostPhys.13.5.110},
}

@article{ce-23a,
  author = "Capizzi, Luca and Eisler, Viktor",
    title = "{Zero-mode entanglement across a conformal defect}",
    eprint = "2303.10425",
    archivePrefix = "arXiv",
    primaryClass = "cond-mat.stat-mech",
    doi = "10.1088/1742-5468/acd68f",
    journal = "J. Stat. Mech.",
    volume = "2305",
    pages = "053109",
    year = "2023"
}

@article{dfms-87,
    author = "Dixon, Lance J. and Friedan, Daniel and Martinec, Emil J. and Shenker, Stephen H.",
    title = "{The Conformal Field Theory of Orbifolds}",
    reportNumber = "EFI-86-42-CHICAGO",
    doi = "10.1016/0550-3213(87)90676-6",
    journal = "Nucl. Phys. B",
    volume = "282",
    pages = "13--73",
    year = "1987"
}

@article{cct-12,
 author = "Calabrese, Pasquale and Cardy, John and Tonni, Erik",
    title = "{Entanglement negativity in quantum field theory}",
    eprint = "1206.3092",
    archivePrefix = "arXiv",
    primaryClass = "cond-mat.stat-mech",
    doi = "10.1103/PhysRevLett.109.130502",
    journal = "Phys. Rev. Lett.",
    volume = "109",
    pages = "130502",
    year = "2012"
}

@article{ac-19,
  author = "Alba, Vincenzo and Calabrese, Pasquale",
    title = "{Quantum information dynamics in multipartite integrable systems}",
    eprint = "1809.09119",
    archivePrefix = "arXiv",
    primaryClass = "cond-mat.stat-mech",
    doi = "10.1209/0295-5075/126/60001",
    journal = "EPL",
    volume = "126",
    number = "6",
    pages = "60001",
    year = "2019"
}

@article{ge-20,
  author = "Gruber, Matthias and Eisler, Viktor",
    title = "{Time evolution of entanglement negativity across a defect}",
    eprint = "2001.06274",
    archivePrefix = "arXiv",
    primaryClass = "cond-mat.stat-mech",
    doi = "10.1088/1751-8121/ab831c",
    journal = "J. Phys. A",
    volume = "53",
    number = "20",
    pages = "205301",
    year = "2020"
}

@article{ep-12,
  title={On entanglement evolution across defects in critical chains},
  author={Eisler, Viktor and Peschel, Ingo},
  journal={Europhysics Letters},
  volume={99},
  number={2},
  pages={20001},
  year={2012},
  publisher={IOP Publishing},
doi="https://doi.org/10.1209/0295-5075/99/20001"
}

@article{pe-12,
  title={Exact results for the entanglement across defects in critical chains},
  author={Peschel, Ingo and Eisler, Viktor},
  journal={J. Phys. A Math. Theor.},
  volume={45},
  number={15},
  pages={155301},
  year={2012},
  publisher={IOP Publishing},
doi="https://doi.org/10.1088/1751-8113/45/15/155301"
}

@article{fc-11,
   author = "Fagotti, Maurizio and Calabrese, Pasquale",
    title = "{Universal parity effects in the entanglement entropy of XX chains with open boundary conditions}",
    eprint = "1010.5796",
    archivePrefix = "arXiv",
    primaryClass = "cond-mat.stat-mech",
    doi = "10.1088/1742-5468/2011/01/P01017",
    journal = "J. Stat. Mech.",
    volume = "1101",
    pages = "P01017",
    year = "2011"
}

@article{Peschel-03,
doi = {10.1088/0305-4470/36/14/101},
url = {https://dx.doi.org/10.1088/0305-4470/36/14/101},
year = {2003},
publisher = {},
volume = {36},
number = {14},
pages = {L205},
author = {Ingo Peschel},
title = {Calculation of reduced density matrices from correlation functions},
journal = {J. Phys. A: Math. Gen.}
}

@article{Peschel-09,
author = "Eisler, Viktor and Peschel, Ingo",
    title = "{Reduced density matrices and entanglement entropy in free lattice models}",
    eprint = "0906.1663",
    archivePrefix = "arXiv",
    primaryClass = "cond-mat.stat-mech",
    doi = "10.1088/1751-8113/42/50/504003",
    journal = "J. Phys. A",
    volume = "42",
    number = "50",
    pages = "504003",
    year = "2009"
}

@article{ssr-17,
  title = {Partial time-reversal transformation and entanglement negativity in fermionic systems},
  author = {Shapourian, Hassan and Shiozaki, Ken and Ryu, Shinsei},
  journal = {Phys. Rev. B},
  volume = {95},
  issue = {16},
  pages = {165101},
  numpages = {18},
  year = {2017},
  publisher = {American Physical Society},
  doi = {10.1103/PhysRevB.95.165101},
  url = {https://link.aps.org/doi/10.1103/PhysRevB.95.165101}
}

@article{rac-16b,
    author = "Ruggiero, Paola and Alba, Vincenzo and Calabrese, Pasquale",
    title = "{Negativity spectrum of one-dimensional conformal field theories}",
    eprint = "1607.02992",
    archivePrefix = "arXiv",
    primaryClass = "cond-mat.stat-mech",
    doi = "10.1103/PhysRevB.94.195121",
    journal = "Phys. Rev. B",
    volume = "94",
    number = "19",
    pages = "195121",
    year = "2016"
}

@article{srrc-19,
    author = "Shapourian, Hassan and Ruggiero, Paola and Ryu, Shinsei and Calabrese, Pasquale",
    title = "{Twisted and untwisted negativity spectrum of free fermions}",
    eprint = "1906.04211",
    archivePrefix = "arXiv",
    primaryClass = "cond-mat.stat-mech",
    doi = "10.21468/SciPostPhys.7.3.037",
    journal = "SciPost Phys.",
    volume = "7",
    number = "3",
    pages = "037",
    year = "2019"
}

@article{sr-19a,
  title = {Entanglement negativity of fermions: Monotonicity, separability criterion, and classification of few-mode states},
  author = {Shapourian, Hassan and Ryu, Shinsei},
  journal = {Phys. Rev. A},
  volume = {99},
  issue = {2},
  pages = {022310},
  numpages = {24},
  year = {2019},
  publisher = {American Physical Society},
  doi = {10.1103/PhysRevA.99.022310},
  url = {https://link.aps.org/doi/10.1103/PhysRevA.99.022310}
}

@article{srgr-18,
    author = "Shiozaki, Ken and Shapourian, Hassan and Gomi, Kiyonori and Ryu, Shinsei",
    title = "{Many-body topological invariants for fermionic short-range entangled topological phases protected by antiunitary symmetries}",
    eprint = "1710.01886",
    archivePrefix = "arXiv",
    primaryClass = "cond-mat.str-el",
    doi = "10.1103/PhysRevB.98.035151",
    journal = "Phys. Rev. B",
    volume = "98",
    number = "3",
    pages = "035151",
    year = "2018"
}

\end{document}